\documentclass{aastex62}

\accepted{for publication in Astronomical Journal}
\shorttitle{Stellar Mass, M/L ratio and color-M/L relation of LSBGs}
\shortauthors{Du et al.}

\begin{document}

\title{Stellar Mass and stellar Mass-to-light ratio$\--$Color relations for Low Surface Brightness Galaxies}

\correspondingauthor{Wei Du, Zheng Zheng}
\email{wdu@nao.cas.cn, zz@nao.cas.cn}

\author{Wei Du}
\affiliation{National Astronomical Observatories, Chinese Academy of Sciences (NAOC), 20A Datun Road, Chaoyang District, Beijing 100012, China}
\affil{Key Laboratory of Optical Astronomy, NAOC, 20A Datun Road, Chaoyang District, Beijing 100012, China}

\author{Cheng Cheng}
\affil{Chinese Academy of Sciences South America Center for Astronomy, National Astronomical Observatories, CAS, Beijing 100101, China}

\author{Zheng Zheng}
\affil{National Astronomical Observatories, Chinese Academy of Sciences (NAOC), 20A Datun Road, Chaoyang District, Beijing 100012, China}
\affil{Chinese Academy of Sciences Key Laboratory of FAST, NAOC, 20A Datun Road, Chaoyang District, Beijing 100012, China}

\author{Hong Wu}
\affil{National Astronomical Observatories, Chinese Academy of Sciences (NAOC), 20A Datun Road, Chaoyang District, Beijing 100012, China}
\affil{Key Laboratory of Optical Astronomy, NAOC, 20A Datun Road, Chaoyang District, Beijing 100012, China}



\begin{abstract}
We estimate the stellar mass for a sample of low surface brightness galaxies (LSBGs) 
by fitting their multiband spectral energy distributions (SEDs) to
the stellar population synthesis (SPS) model. 
The derived stellar masses (log $M_{*}/M_{\odot}$) span from 7.1 to 11.1, 
with a mean of log $M_{*}/M_{\odot}$=8.5, which is lower than that for normal galaxies.
The stellar mass-to-light ratio ($\gamma^{*}$) in each band varies little with the absolute magnitude,
but increases with higher $M_{*}$. This trend of $\gamma^{*}$ with $M_{*}$ is even stronger in bluer bands.  
In addition, the $\gamma^{*}$ for our LSBGs slightly declines 
from $r$ band to the longer wavelength bands.
The log $\gamma_{*}^{j}$($j$=$g$,$r$,$i$,and $z$) 
have relatively tight relations
with optical colors of $g-r$ and $g-i$.
Compared with several representative $\gamma^{*}$-color relations (MLCRs)
from literature, our MLCRs based on LSBG data are consistently 
among those literature MLCRs previously defined on diverse galaxy samples,
and the existing minor differences between the MLCRs
are more caused by the differences in the SED model ingredients 
including initial mass function, star formation history,
and stellar population model, and the line fitting techniques,
galaxy samples, and photometric zero-point as well, rather than the
galaxy surface brightness itself which distinguishes LSBGs from HSBGs.  
Our LSBGs would be very likely to follow those
representative MLCRs previously defined on diverse galaxy populations,
if those main ingredients were taken into account.


\end{abstract}

\keywords{}


\section{Introduction} \label{sec:intro}
Galaxies with central surface brightnesses fainter than the night sky ($\sim$ 22.5 $B$ mag arcsec$^{-2}$ ) 
are defined as Low Surface Brightness Galaxies (LSBGs;
\citep[e.g.][]{Impey1997,Impey2001,Ceccarelli2012}. 
In the local universe, LSBGs take up a fraction of $\sim$30\%--60\% in number 
\citep[e.g.][]{McGaugh1995,McGaugh1996,Bothun1997,O'Neil2000,Trachternach2006,Haberzettl2007} 
and $\sim$ 20\% in dynamical mass \citep[e.g.][]{Minchin2004} among all the galaxies.
Generally speaking, LSBGs are abundant in H{\sc{i}} gas 
but deficient in metal ($\leq 1/3$ solar abundance) and 
dust \citep[e.g.][]{McGaugh1994,Matthews2001},
and they have fairly low star formation rates (SFR) 
\citep[e.g.,][]{Das2009,Galaz2011,Lei2018}, 
which are evident that only a small number of H{\sc{ii}} regions inhabit in their diffuse disks. 
They also have lower stellar mass densities, 
comparing to their High Surface Brightness Galaxy (HSBG) counterparts (normal galaxies) 
\citep[e.g.][]{de Blok1996,Burkholder2001,O'Neil2004,Trachternach2006}. 
These special properties imply that
LSBGs could have different formation and evolutionary histories from normal galaxies\citep[e.g.][]{Huang2012}. 

Galaxy stellar mass, $M_*$, is a critical physical property 
for studying galaxy formation and evolution, 
because its growth is directly related to galaxy formation and evolution.
A widely used simple method to estimate $M_*$ is 
to multiply the measured galaxy luminosity, $L$,  
with a fixed stellar mass-to-light ratio ($\gamma^{*}$). 
However, different stellar populations have very different spectral energy distributions
with younger stars dominating bluer bands and older stars dominating redder bands. 
Therefore, the $\gamma^{*}$s in different bands need to be calibrated separately for different populations 
and different photometric bands.
In this case, the technique of modeling the broadband photometry (SED-fitting) to stellar population synthesis (SPS) models
is used to estimate the stellar mass of the galaxy.
Usually, the stellar masses for
“normal” galaxies (i.e., not including pathological SFHs) can be recovered at the $\sim$ 0.3 dex level 
(1$\sigma$ uncertainty) by the broadband SED fitting. 
This uncertainty does not include potential systematics in the underlying SPS
models \citep{Conroy2013}.
Then, another convenient technique to derive the galaxy stellar mass is
to use the relation between color and stellar mass-to-light ratio of the galaxies.
So far, the $\gamma^{*}$s in different photometric broad bands
have been derived as a function of galaxy colors by various studies
\citep[e.g.,][]{Bell2001,Bell2003,Portinari2004,Zibetti2009,McGaugh2014}.
However, the existing prescriptions are mostly calibrated 
for normal galaxies, which have very different properties from LSBGs. 
The star-forming main sequence (log $SFR$ - log $M_*$) of dwarf LSBGs has a steep slope of approximately unity,
distinct from the shallower slope of more massive spirals \citep{McGaugh2017}.
 Generally, a slope of unity would agree with galaxies forming early in the universe 
 and subsequently forming stars at a nearly constant specific star formation rate (sSFR).  
 In contrast, a shallow slope implies that low-mass galaxies were formed 
 recently in a shorter time scale with a higher sSFR 
 (sSFR is roughly inversely proportional to the age of stellar disk).  
These different formation scenarios could potentially lead to different $\gamma^{*}$s.
Therefore, we estimate the $\gamma^{*}$s and stellar masses of a sample of over 1,000~LSBGs defined 
in our previous work of \citet{Du2015,Du2019}, by fitting their multi-band (from UV to NIR) spectral energy distributions (SEDs) 
to the stellar population models, and investigate the correlations between the $\gamma^{*}$ and observed colors for the LSBG sample.    

We briefly introduce the LSBG sample, 
and show the multi-wavelength photometric band data in \S 2. 
We then show data reduction and photometry in \S 3.
We demonstrate the multi-band SED fitting process and 
show the derived LSBG stellar mass distribution in \S 4.
We explore the distributions of the derived LSBG M/Ls and their correlations 
with galaxy colors in \S 5 and discuss the results in \S 6.
Throughout this paper, the distances we used to convert apparent magnitude
to absolute magnitude and luminosity 
are from the ALFALFA catalog \citep{Haynes2018}, 
which adopted the Hubble constant of H$_{0}$=70 km~s$^{-1}$~Mpc$^{-1}$. 
Magnitudes in this paper are all in the AB magnitude system. 

\section{Sample and data} \label{sec:data}
\subsection{LSBG Sample}\label{subsec:sample}
We have selected a sample of 1129 LSBGs from the 
combination of the $\alpha$.40 H{\sc{i}} survey  \citep{Haynes2011} 
and the SDSS DR7 photometric survey \citep{Abazajian2009}.
We briefly introduce the sample selection and properties below,
and details and related studies could be found in \citet{Du2015} and \citet{Du2019}.

LSBGs are very sensitive to the sky background, 
so it is crucial to precisely subtract the sky background 
from the galaxy image before photometry. 
Unfortunately, the sky backgrounds have been overestimated by the SDSS photometric pipeline 
for galaxy images in their $ugriz$ bands,
which consequently results in an average underestimation of $\sim$0.2 mag in luminosity of bright galaxies  
\citep{Lauer2007,Liu2008,Hyde2009,He2013}
and $\sim$0.5 mag of LSBGs \citep{Lisker2007}.  
To improve the sky subtraction, we re-estimate the sky background
of $g$- and $r$-band images for each galaxy in the $\alpha.$40-SDSS DR7 sample,
using a fully tested method of sky estimation\citep{Zheng1999}.
The method fits all the sky pixels on the object-masked image 
row-by-row and column-by-column, and 
is designed for a better estimate of the sky background map
for galaxies with faint outskirts \citep{Zheng1999,Wu2002}
and LSBGs \citep{Du2015}.
More details about this sky-background-estimation method
and its applications on bright galaxies and LSBGs have
been reported in  \citet{Zheng1999},
\citet{Wu2002}, and \citet{Du2015}.
On sky-subtracted images in $g$ and $r$ bands,
we measure the magnitudes of galaxies using the SExtractor code \citep{Bertin1996},
and fit the radial surface brightness profiles of galaxies to exponential profile models
using the Galfit code \citep{Peng2002}.
Then, the magnitude from SExtractor and the results from Galfit including disk scale length, $r_{s}$, and 
minor-to-major axial ratio, $b/a$, are used
to calculate the disk central surface brightness in $g$ and $r$ bands, 
which are then combined to be converted to the disk central surface brightness in $B$ band, $\mu_{0}$(B),
according to the transformation formula of \citet{Smith2002}.
Finally, based on the traditional definition for LSBGs ,
we select 1129 non-edge-on galaxies (b/a $>$ 0.3 in both $g$ and $r$ bands)
with $\mu_{0}(B) \geq$  22.5 mag arcsec$^{-2}$ from the entire $\alpha.$40-SDSS DR7 sample (12,423 galaxies)
to form our LSBG sample.

The H{\sc{i}}-selected LSBG sample, inhabiting in low density
environment \citep{Du2015} and dominated by dwarf galaxies 
in luminosity and late-type galaxies in morphology, 
has extended the parameter space 
covered by the existing LSBG samples
to fainter luminosity, lower H{\sc{i}} gas mass, and bluer color \citep{Du2019}. 
More details are available in \citet{Du2015} and \citet{Du2019}.

\subsection{Photometric data}
We collected the available scientific images for
each galaxy of our LSBG sample in the passbands of FUV, NUV ($GALEX$; \citet[]{Martin2005})
$ugriz$ (SDSS DR7; \citet[]{York2000}), 
and $YJHK$ (UKIDSS LAS DR10; \citet[]{Lawrence2007}).
We note that the sky areas covered by different surveys
do not completely overlap each other 
due to different survey strategies. 
Therefore, not all the galaxies in our LSBG sample are available 
in all the 11 passbands. We list the number of galaxies which are available in each of the 11 passbands in Table ~\ref{tab:multibands}. 
All the LSBGs in our sample have been observed by the SDSS DR7.
However, combining the optical with UV bands, 924 galaxies are available in both SDSS DR7 and GALEX DR6 surveys.
Combining the optical with NIR bands, 672 galaxies have been observed by both SDSS DR7 and UKIDSS LAS DR10 surveys.
Furthermore, combining the optical, UV and NIR, 
only 544 galaxies are available in all the 11 bands.  
\begin{table}
\begin{center}
\caption{The number of LSBG scientific images available in multi-passbands. }
\label{tab:multibands}

\begin{tabular}{lc}
\hline
 Passband & number\\
\hline
 GALEX FUV & 924  \\
 GALEX NUV & 932 \\
 SDSS u & 1129\\
 SDSS g & 1129\\
 SDSS r & 1129\\
SDSS i & 1129\\
SDSS z & 1129\\
UKIDSS Y & 717\\
UKIDSS J & 697\\
UKIDSS H & 709\\
UKIDSS K & 717\\
   \hline
  \end{tabular}
   \end{center} 
 \end{table}

\subsection{Multi-band photometry}\label{sec:phot}
\subsubsection{Image reduction}
The scientific images for galaxies in our LSBG sample
are all bias subtracted, dark subtracted (only for NIR frames), 
flat-fielded and flux-calibrated by the survey teams. 
We just need to start the data reduction
from sky subtraction. 

As for the UV sky background, 
we directly use the sky background map provided by the GALEX team. 
However, for the optical and NIR band,
only an average sky value is provided for a galaxy image
by SDSS DR7 or UKIDSS LAS DR10,
not to mention the problem of overestimation of sky background by SDSS DR7 photometric
pipeline (see \S ~\ref{sec:data}). 
So we estimate the sky background map for LSBG image in each of the $ugriz$ and
$YJHK$ bands, using the row-by-row and column-by-column fitting method \citep{Zheng1999}
on the image with all the detected objects being masked out.
As a fully tested sky estimation method, 
it has been successfully practiced for bright galaxies
with faint extended outskirts \citep{Zheng1999,Wu2002}
and for LSBGs \citep{Du2015},
and the details are elaborated in \S 3.1 in \citet{Du2015}.

\begin{figure}[ht!]
\centering
\includegraphics[width=0.8\textwidth]{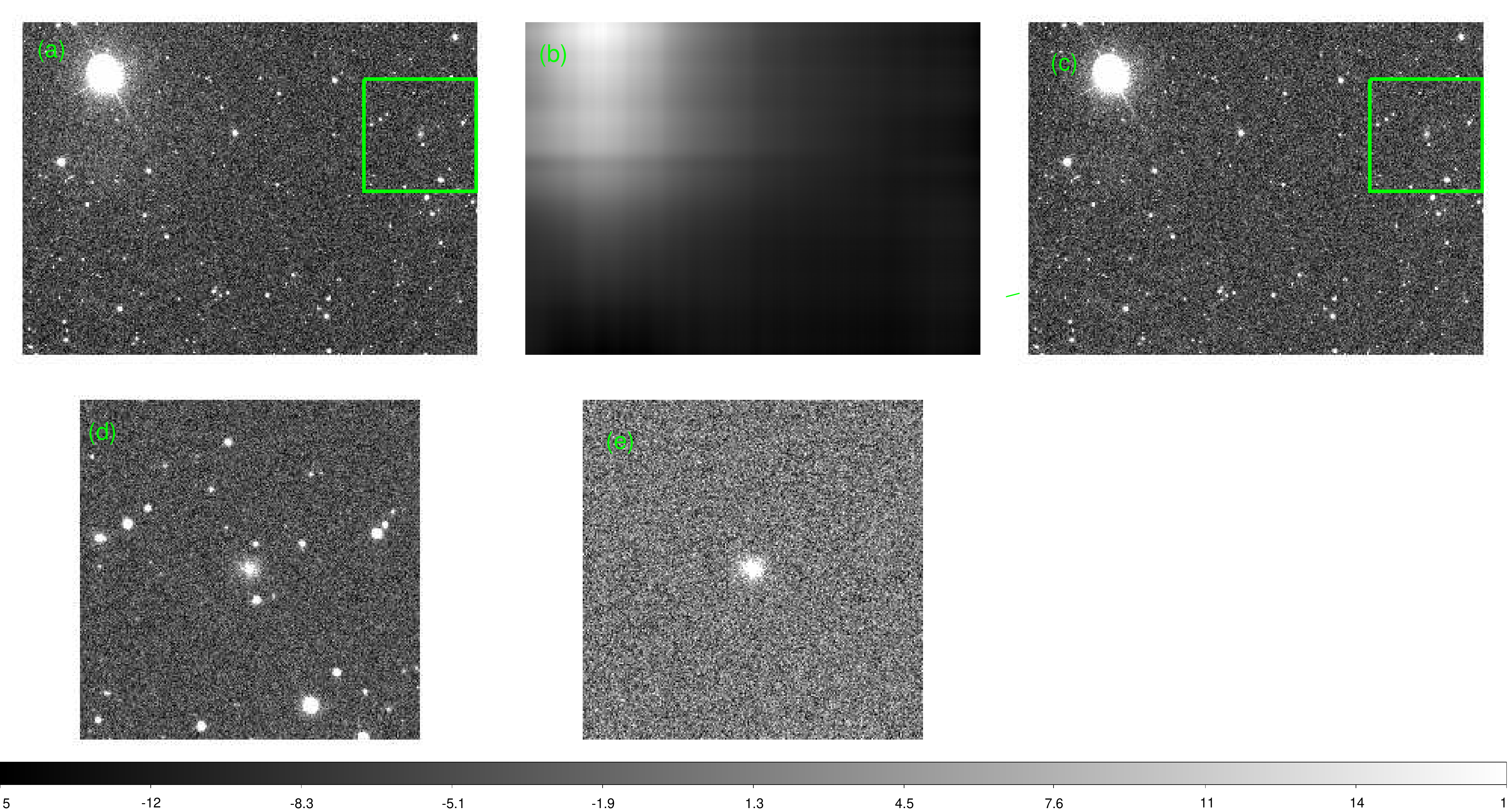}
\caption{Reduction of the SDSS $r$-band image of an example LSBG galaxy
AGC 192669. Panel (a) shows
the original $r$-band scientific image frame from SDSS DR7, 
with AGC 192669 centering at the green square region of size 200~arcsec $\times$ 200~arcsec. 
Panel (b) displays the sky background map we constructed by using the row-by-row
and column-by-column fitting method \citep{Zheng1999}. 
Panel (c) shows the sky-subtracted
image, which is produced by subtracting the sky background map (Panel (b))
from the original galaxy image (Panel (a)). 
Panel (d) shows the region within the green square of the size 200$\arcsec \times$ 200$\arcsec$,
centering at the center of the galaxy AGC 192669. 
On the trimmed image,  
we masked all other objects (except for the central target
galaxy itself) detected by SExtractor, and generated the image in Panel (e)
by replacing the pixel values within the masked regions by the average background 
value. The image shown in panel (e) is for further photometric measurements}.\label{fig:general}
\end{figure}

After subtracting the sky background
for the LSBG image, 
we use SExtractor \citep{Bertin1996} to make a systematic, homogeneous
photometric measurement on the LSBG images of all the 11 bands,
by performing the pixel-to-pixel photometry in dual-image mode. 
In the dual-image mode,
the galaxy image in the $r$ band is used as a reference
for source detection and 
photometric aperture definition (center position, size and shape), 
and then images in all other passbands for the galaxy 
are photometrically measured within 
the same aperture as defined by the reference image. 
So, the dual-image mode of SExtractor requires that
galaxy images in all passbands must match with the reference image
in dimension, orientation, pixel scale, image size, and object position. 
We use the $r$-band galaxy image with sky subtracted as the reference,
so we have to resample the sky-subtracted images of the galaxy in all other bands 
(FUV NUV, $ugiz$, $YJHK$) using cubic interpolations to match with the reference image
in image orientation, pixel scale, image size, and object position. 
The matched images in all bands are then trimmed to have a same size of 200~arcsec $\times$ 200~arcsec,
åwith the target galaxy at the center of the trimmed image. 
Then, we keep the centering galaxy (target) on each trimmed image, and
mask all the other detected objects (by SExtractor with 3$\sigma$ as the
minimum detection threshold) from the trimmed image
to avoid the light contamination from adjacent objects.
We then fill the masked regions with average values
of the background pixel of each object-masked image.
Such reduced images are ready for further photometry
in \S ~\ref{sec:photometry}.
For a better illustration of the image reduction process, 
we take the $r$-band image of the galaxy AGC 192669 
in our LSBG sample as an example in Figure ~\ref{fig:general},
where panels (a) and (c), respectively show the original and 
sky-subtracted image frames. In order to more clearly demonstrate
the quality of our sky subtraction, 
we compare the distribution of the background pixel values of the sky-subtracted image 
(panel (c)) with the distribution of the background pixel values of the original image (panel (a)) 
with a simple mean background value subtracted from in Figure 2, 
where the image background is closer to zero after our sky subtraction (black line).


\subsubsection{Photometry}\label{sec:photometry}
We use SExtractor to define the position and aperture of 
the galaxy in the $r$-band image, and then used the $r$-band
position and aperture information to measure the galaxy magnitude
within each filter of $FUV$, $NUV$, $u$, $g$, $r$, $i$, $z$, $Y$, $J$,
$H$, and $K$ (using the SExtractor dual image mode).
As the aperture definitions do not vary between wavebands,
this measurement gives internally consistent colors.
The measured magnitudes in all bands are corrected for Galactic extinction
using the prescription of \citet{Schlafly2011}.
We show the the aperture of the galaxy AGC 192669
on reduced image in each band in Figure ~\ref{fig:phot} for example. 

This aperture is the automatic aperture (AUTO), 
as among the various magnitude types that SExtractor provides
(isophotal, corrected isophotal, fixed-aperture, 
AUTO, and Petrosian),
the automatic aperture (AUTO), inspired by Kron's ``first moment'' algorithm (see details in \citet{Kron1980}),
is a flexible and accurate elliptical aperture whose elongation $\epsilon$ and position angle $\theta$ are 
defined by the second order moments of the object's light distribution. 
Within this aperture, the characteristic radius $r_{1}$ is 
defined as $r_{1}$= $\frac{\sum rI(r)}{\sum I(r)}$,
 which is weighted by the light distribution function.  
\citet{Kron1980} and \citet{Infante1987} have verified that, for stars and galaxy profiles convolved 
with a Gaussian, more than 90$\%$ of the flux is expected to lie within a circular aperture 
of radius $kr_{1}$ if k=2, almost independently of their magnitudes. 
This changes if an ellipse 
with $\epsilon kr_{1}$ and $\frac{kr_{1}}{ \epsilon}$ is considered as the principal axes. 
By choosing a larger k=2.5, more than 96$\%$ of the flux is captured within the elliptical aperture. 
So, the AUTO magnitudes are intended to give the more precise estimate of ``total magnitudes", 
at least for galaxies.  
More details about the AUTO photometry
could be found in SExtractor manual and \citet{Kron1980},\citet{Infante1987},
and \citet{Bertin1996}.
During our measurments, we keep k=2.5, the recommended setting 
by SExtractor \citep{Bertin1996}, 
and our following studies would be based on 
the AUTO magnitude in AB magnitude system.

\section{Stellar Mass}\label{sec:mstar}
Galaxies emit electromagnetic radiation over the full possible wavelength
range, and the distribution of energy over wavelength is called
the Spectral Energy Distribution (SED) which is our primary source
of information about the properties of the unresolved galaxy. In general,
the different physical processes occurring in galaxies all leave
their imprints on the global and detailed shapes of the spectra or
SEDs. Therefore, we can constrain the galaxy stellar mass
by fitting models to its SED.
In \S ~\ref{sec:photometry}, we derived the multi-band magnitudes
of each LSBG in our sample, so we can construct the SED for each LSBG.
Since the UV light comes from regions where hot and
young stars reside and the NIR light is the best tracer
of old stars which dominate the stellar mass of a galaxy, 
our SEDs covering multi-bands
from UV to NIR allow us to include contribution of both young and old stars
to stellar mass.
It should be noted that for the LSBGs, 
which have not been observed in all the 11 bands
we only use the available bands to construct the SED.

\subsection{SED-fitting}
MAGPHYS \citep{da Cunha2008}
is one of the widely used SED-fitting codes \citep[e.g.,][]{Zheng2015},
which uses the 2007 version of \citet{Bruzual2003} stellar population synthesis (SPS) model (CB07) 
covering a wavelength
range from 91~\AA~to 160 $\mu$m, different ages from 0.1 Myr
to 20 Gyr and various metallicities (Z) from 0.02 to 2 times solar.
The star formation history (SFH) is described
by an underlying continuous model (exponentially declining star
formation)  with instantaneous bursts superimposed. 
The initial mass function (IMF) of Chabrier \citep{Chabrier2003}
is assumed,
and a simple two-component
dust model of \citet{Charlot2000} is adopted to describe the attenuation of
the stellar light by the dust.
Using MAGPHYS\citep{da Cunha2008},
we fit the SPS models to the multi-band SED 
of each LSBG in our sample to estimate the galaxy stellar mass.  
For example, we show the best-fitting
model and residual by using MAGPHYS
for the LSBG, AGC 192669 in the top two panels in Figure ~\ref{fig:sed_fitting}.
We note that MAGPHYS gives both the stellar mass of the best-fitting model
and the stellar mass distribution.
In this paper, we use the mean value of stellar mass 
from the given stellar mass distribution
as our derived galaxy stellar mass, $M_{*}$.
  
Checking the fitting results from MAGPHYS, 
we found that MAGPHYS always give the lower limit of the model in stellar mass 
for 77 galaxies out of the total LSBG sample (1129 galaxies), 
so we would exclude these 77 galaxies
in subsequent analysis,
and the sample for further investigations
include 1,052 LSBGs (Research Sample - R sample).
In order to check whether the R sample (1040 LSBGs) 
is representative of the total LSBG sample (T sample; 1129 LSBGs),
we compare them in terms of 
physical property in Figure ~\ref{fig:subsample}.
Compared with the T sample (black),
the R sample (green) has quite similar distributions
in the main properties of magnitude, color, 
central surface brightness, size,
and color versus H{\sc{i}} mass 
except for that the R sample lack the very low-redshift, faint,
and H{\sc{i}}-poor galaxies.
Therefore, the R sample is a good representative of our total LSBG sample,
and our further analysis in the paper would be based on the R sample, 
which may still be named the LSBG sample later.

\subsection{Stelar mass distribution}
The derived stellar masses using MAGPHYS 
are shown in Figure ~\ref{fig:mstar_distri} for the R sample,
ranging from log$M_{*}/M_{\odot}\sim$ 7.1  to 11.1, 
with a mean log$M_{*}/M_{\odot}$=8.47 and a median log$M_{*}/M_{\odot}$=8.48
for the R LSBG sample. 
which is considerably lower than the stellar mass for the normal galaxies. 
Furthermore, LSBGs which have lower surface brightnesses
(fainter than 25.0 $mag/arcsec^{2}$) tend to have lower stellar mass, 
and LSBGs which have higher stellar masses tend to have 
higher surface brightness (the right panel of Figure ~\ref{fig:mstar_distri}).
 
\section{Stellar Mass-to-light ratio}\label{sec:m2l}
The best approach to measure stellar mass-to-light ratio, $\gamma^{*}$,
is to fit SEDs simultaneously in multi-passbands, with at least one in the NIR
to break the age-metallicity degeneracy. 
We show the $\gamma^{*}$ derived from fitting the UV-optical-NIR SEDs
of our galaxies in Figure ~\ref{fig:m2l_distr}.
The UV is strongly affected by the young, luminous, blue stars 
formed in recent star formation history (SFH) of a galaxy.
These stars produce a large amount of the UV light
and contribute most to the fluxes of galaxies,
so having the UV band involved into the galaxy SED fitting
should provide stronger constraints on the SFH, average age, metallicity, stellar mass, and $\gamma^{*}$
of galaxies via the luminosity-weighted SED fitting.
Compared to the average $\gamma^{*}$ measured in optical and NIR bands, 
the $\gamma^{*}$ in UV bands suffer more perturbations 
because those young, luminous, blue stars contribute little to the stellar mass of a galaxy
 \citep{McGaugh2014}, so it is not informative to investigate the $\gamma^{*}$ in UV bands derived from SED-fitting.
Here, we only show the log $\gamma^{*}$ measured in optical $ugriz$ and NIR $YJHK$ bands for the R LSBG sample
in Figure ~\ref{fig:m2l_distr} (a) $\sim$ (i),
and the log $\gamma^{*}$ ($\gamma^{*}$) are
-0.48 (0.33), -0.40 (0.40), -0.33 (0.47), -0.40 (0.40), -0.39 (0.41), -0.40 (0.40), -0.46 (0.35), -0.55 (0.28), and -0.66 (0.22)
in $ugrizYJHK$ bands, respectively,
which are lower than the $\gamma^{*}$ for normal galaxies.
In Figure ~\ref{fig:m2l_distr} (j), the mean $\gamma^{*}$ 
slightly declines as the wavelength band moves from $r$ to $K$.
Such a declining trend for $\gamma^{*}$
of the normal disk galaxies from $V$ to $[3.6]$ through $I$ and $K$ 
has been reported to be stronger \citep{McGaugh2014}. 
According to Figure 1 in \citet{Wilkins2013},
the $\gamma^{*}$ measured at different wavelengths of a stellar population
are dependent on the specific SED shape of the population,
with a lower $\gamma^{*}$ at the wavelength with a higher specific SED flux.
So, the slight declining trend of $\gamma^{*}$ of our sample
from $r$ to $K$ implies that the overall LSBGs 
of our sample have a slightly rising SED shape from $r$ to NIR.

In Figure ~\ref{fig:m2l},the $\gamma^{*}$ measured in each band of $ugrizYJHK$ 
is shown against the absolute magnitude (left column)
and stellar mass (right column) for the R LSBG sample.
While the distribution of $\gamma^{*}$ in $z$ band
for a sample of galaxies drawn from SDSS main galaxy sample 
(bright galaxies with Petrosian $r$-band magnitudes in the range of 14.5 $< r <$ 17.7 mag)
is strongly dependent on galaxy luminosity (see Fig. 13 of \citet[]{Kauffmann2003}),
it shows that $\gamma^{*}$ for the bright galaxies (e.g., brighter than around -13 mag in $r$ band)
of our LSBG sample slowly changes with the absolute magnitude 
in any band from $u$ to $K$ (left column), as the slopes of the fitting lines
are all around zero (within $\pm$0.04) if we forced only the bright galaxies (with absolute magnitude in the corresponding band
brighter than -16 mag) to be fitted by a linear line in each panel
of the left column.
However, $\gamma^{*}$ for the R LSBG sample slightly increases with the galaxy stellar mass (right column),
and this increasing trend (represented by the black dotted line which is a linear fitting for all the R LSBG sample galaxies)
is stronger in shorter/bluer wavelength bands than longer/redder bands.
This is quantitatively evidenced by the steepest slope of the fitting line
in $u$ (top panel) and nearly flat slope of the fitting line  in $K$ band (bottom panel)
in Figure ~\ref{fig:m2l}. 
Furthermore, the scatter of data points in $\gamma^{*}$ in any band
is becoming narrower at higher stellar mass (right column),
indicating less diversity in star formation history (SFH).

\section{Color v.s. stellar mass-to-light ratio}\label{sec:ms}
 \subsection{Color - $\gamma^{*}$ relation for our LSBGs}
Various prescriptions for predicting $\gamma^{*}$ 
from the observed colors have been calibrated for normal galaxies
previously \citep[e.g.,][]{Bell2001,Bell2003,Portinari2004,Zibetti2009,McGaugh2014}.
In contrast, our correlation is based on a sample of LSBGs,
which is dominated by dwarf LSBGs (Figure 1 in \citet{Du2019})
with $\sim$ 50$\%$ of the R sample fainter than $r$= 17.5~mag (Figure 2(b) in \citet{Du2019}),
almost all fainter than 21~mag~arcsec$^{-2}$ in the $r$ surface brightnesses, $\mu_{r}$,  
(Figure 2(d) in \citet{Du2019}), and 73$\%$ bluer than $g$-$r$=0.4~mag (Figure ~\ref{fig:subsample}(b)).  
As the dwarf LSBGs have been reported to form a distinct sequence from more massive normal galaxies in
the star-forming main sequence \citep{McGaugh2017}, 
we expect to study the $\gamma^{*}$-color relation (MLCR) based on our LSBG sample, 
which has a large fraction of dwarf LSBGs.

We fit the relations of $\gamma^{*}$ measured in each of the $grizJHK$ bands
to the optical colors ($u-g$, $u-r$, $u-i$, $u-z$, $g-r$, $g-i$, $g-z$, $r-i$, and $r-z$)
and the NIR colors ($J-H$, $J-K$, and $H-K$), respectively,
for the LSBG sample, in the form of
log($\gamma^{*}$) = $a_{\lambda}$ + $b_{\lambda}\times$color.
The fitting method is the bi-square weighted line fitting method, 
which is the same as the fitting method that was used by \citet{Bell2003} MLCRs. 
To test the goodness of the fit to our data, we calculate and show the Pearson correlation
coefficient (PCC) for each of these fits in Table ~\ref{tab:Pearson}.
The PCC is a measure of the linear correlation between two variables,
and it has a value between +1 and -1, 
where +1(-1) is a total positive (negative) linear correlation, 
and 0 is no linear correlation.
As shown in Table ~\ref{tab:Pearson}, 
using $u$-colors as a $\gamma^{*}$ estimator for our sample
results in low PCCs ($<$ 0.5). This is presumably because
of two reasons. One reason is that, compared with relatively redder bands ($g$ and $r$),
 $u$ band is more affected
by the recently formed young stars, which would
cause considerable perturbations to the average $\gamma^{*}$
in $u$ band than $g$ and $r$ bands. Another reason might
be due to the quality of SDSS $u$-band images which are reported
to have scattered light problems that may cause relatively larger errors
in the $u$ band fluxes than other SDSS bands
and then perturb $u$-colors (see the `Caveats' in SDSS websites).
In this case, the $u$-colors do not seem to be good estimators of $\gamma^{*}$
for our sample. 

\citet{McGaugh2014} found
that the solar metallicity model from \citet{Schombert2009} changes in color
as it ages from 1 to 12 Gyr by $\Delta$(B-V)=0.37 and $\Delta$(J-K)=0.03,
demonstrating that NIR colors (such as $J-H$, $J-K$ and $H-K$ here) are
much less sensitive indicators of $\gamma^{*}$ than optical colors. Thus, 
it is expected that using NIR colors as a $\gamma^{*}$ estimator
results in nearly zero PCCs in Table ~\ref{tab:Pearson}, 
and $g-r$ and $g-i$ colors, with the greatest PCCs ( mostly$>$0.5) are instead
more sensitive indicators of $\gamma^{*}$ in $g$, $r$, $i$ and $z$ bands
for our sample. The PCCs are declining as the wavelength band goes redder.
This is because the variation of $\gamma^{*}$ with color is expected 
to be minimized in the NIR, so the $\gamma^{*}$ in a NIR band
is almost a constant, being independent on colors.
In this case, we would only focus on 
investigations of independently using $g-r$ and $g-i$ colors as 
estimators of $\gamma^{*}$ measured in $griz$ bands.
We show ``robust'' bi-square weighted line fits (black solid lines) 
of log $\gamma_{*}^{j}$ ($j$=$g$,$r$,$i$, and $z$)
with $g$ - $r$ color in the left column in Figure ~\ref{fig:m2l_color_gr} 
and with $g$ - $i$ color in the left column in Figure ~\ref{fig:m2l_color_gi}.
The coefficients of the bi-square weighted line fitting relations are tabulated in Table ~\ref{tab:m2l_color},
which provides a direct comparison with tables in other published papaers,
such as Table 7 of
\citet{Bell2003} and Table B1 of \citet{Zibetti2009}.
The detailed comparison will be presented 
in \S ~\ref{sec:com}.

 \begin{table*}
\caption{Pearson correlation coefficients for the MLCRs.}
\label{tab:Pearson}
\begin{center}
\begin{tabular}{lccccccc}
\hline
\hline
 color & PCC$_{g}$ & PCC$_{r}$  &PCC$_{i}$ & PCC$_{z}$ &PCC$_{J}$ & PCC$_{H}$ &PCC$_{K}$ \\
\hline
\hline
 $u-g$&  0.02&   0.05&   0.01&   0.24&    0.00&  0.04&   0.06\\
$u-r$&  0.25&   0.24&   0.20&   0.40&    0.20&  0.21&   0.09\\
$u-i$&  0.26&   0.28&   0.19&   0.47&    0.21&  0.22&   0.13\\
$u-z$&  0.48&   0.47&   0.49&   0.21&    0.38&  0.37&   0.30\\
$g-r$&  0.65&   0.49&   0.56&   0.39&    0.49&  0.45&   0.36\\
$g-i$&  0.67&   0.61&   0.48&   0.59&    0.51&  0.42&   0.37\\
$g-z$&  0.49&   0.46&   0.52&   0.02&    0.43&  0.38&   0.39\\
$r-i$&  0.11&   0.17&   0.03&   0.29&    0.07&  0.06&   0.04\\
$r-z$&  0.28&   0.31&   0.34&   0.16&    0.22&  0.25&   0.29\\
$J-H$&  0.09&   0.04&   0.03&   0.02&    0.29&  0.17&   0.08\\
$J-K$&  0.00&   0.06&   0.06&   0.02&    0.16&  0.03&   0.27\\
$H-K$&  0.09&   0.10&   0.11&   0.02&    0.07&  0.11&   0.30\\

 \hline
 \hline
  \hline
  \end{tabular}
   \end{center} 
 \end{table*}

\begin{table*}
\caption{The fitting parameters for the MLCRs in the form of log($\gamma^{*}$)=$a_{\lambda}$+($b_{\lambda}\times$color).}
\label{tab:m2l_color}
\begin{center}
\begin{tabular}{lcccccccccccccc}
\hline
\hline
color & $a_{g}$ & $b_{g}$  &$a_{r}$ & $b_{r}$ &$a_{i}$ & $b_{i}$ &$a_{z}$ & $b_{z}$\\
\hline
\hline
$g-r$&-0.857&1.558&-0.700&1.252&-0.756&1.226&-0.731&1.128\\
$g-i$&-1.152&1.328&-0.947&1.088&-0.993&1.057&-0.967&0.996\\
 \hline
 \hline
  \hline
  \end{tabular}
   \end{center} 
 \end{table*} 
 
\subsection{Comparison with other color - $\gamma^{*}$ relations}\label{sec:com}
The stellar mass-to-light ratios of galaxies in this work are derived 
by fitting stellar population synthesis (SPS) models to the observed SEDs,
so the derived stellar mass-to-light ratio (hence the stellar mass-to-light ratio -- color relation - MLCR) 
should be dependent on SPS models,  
since different SPS models do not usually take the same prescription for 
the primordial ingredients, including the assumption of initial mass function (IMF)
, the stellar evolution theory in the form of isochrones, 
the treatment of star formation history (SFH), metallicity distribution, and
TP-AGB phase.
Therefore, we compare our MLCRs
with several other representative MLCRs
of \citet[hereafter B03]{Bell2003},
\citet[hereafter Z09]{Zibetti2009},
\citet[hereafter IP13]{Into2013},
\citet{Roediger2015} based on BC03 model (hereafter RC15(BC03))
and FSPS model (hereafter RC15(FSPS)),
and \citet[hereafter H16]{Herrmann2016}. 
Since our relations (and those of Z09, RC15, and H16) are based on
a \citet{Chabrier2003} stellar IMF, while B03 is based on a `diet' Salpeter IMF
and IP13 is based on a \citet{Kroupa1998} IMF, we have applied zero-point offsets
for MLCRs of B03 and IP13 for a better comparison. 
B03 noted that log$\gamma_{*}$ 
should be added by 0.15, 0.0, -0.1, -0.15, -0.15, -0.15,
-0.35 dex to be converted from their `diet' Salpeter IMF
to the \citet{Salpeter1955}, \citet{Gould1997}, \citet{Scalo1986},
\citet{Kroupa1993}, \citet{Kroupa2002},
\citet{Kennicutte1983}, or Bottema 63$\%$ maximal IMFs.
 We follow \citet{Gallazzi2008} and \citet{Zibetti2009} to reduce B03-predicted log$\gamma_{*}$
 by -0.093 dex to convert from the `diet' Salpeter IMF to a \citet{Chabrier2003} IMF.
Following H16, we added 0.057 dex to the IP13-predicted log$\gamma_{*}$ which is based on a Kroupa IMF 
to adjust to a Chabrier IMF.

With all the literature MLCRs adjusted (if needed) to a Chabrier IMF,  
we overplot the relations of log $\gamma_{*}^{j}$ ($j$=$g$,$r$,$i$, and $z$)
with $g$ - $r$ color in the left column in Figure ~\ref{fig:m2l_color_gr}
and with $g$ - $i$ color in the left column in Figure ~\ref{fig:m2l_color_gi}
by different colored dashed lines.
In comparison to each other,
the B03 MLCR (red line) has the shallowest slope
and the Z09 MLCR (blue line) has the steepest slope 
of all the colored lines in each panel,
and our linear fitting line (black solid line) for the LSBG data
is among these colored literature MLCRs in each panel.
B03 MLCRs (red dashed lines) do not appear to fit our LSBG data well
and they overestimate the stellar mass-to-light ratios
for our LSBGs in all panels. 
In quantity, we show the distributions of both B03 MLCR-based (red) 
and our MLCR-based (black) $\gamma_{*}$ (red) for our LSBGs in Figure ~\ref{fig:offset_Bell},
and roughly derive a systematic offset $\Delta$log$\gamma_{*}^{j}$= 0.26~dex for $j$=$g$ and $r$ 
(predicted from $g-r$ color) between BC03 and our MLCRs.
IP13 (green dashed lines) and RC15(FSPS) (grey dashed lines) MLCRs
are slightly higher in all panels whereas Z09 (blue dashed lines) and RC15(BC03) (cyan dashed lines) MLCRs
fit our LSBG data well and match our fitting lines (except for being a little steeper mostly).

Such differences between the MLCRs from various studies
are probably due to the variety of galaxy samples, linear fitting technique,
and SED model ingredients mainly 
consisting of SP model, IMF, and SFH.
All of these factors would be discussed in $\S$~\ref{sec:discuss}. 

 \section{Discussion for variance between MLCRs}\label{sec:discuss}
\subsection{ Variety of galaxy samples}
 B03 MLCR is based on a sample of mostly bright (13 $\leq r \leq$17.5~mag)
 and high surface brightness (HSB; $\mu_{r} <$21~mag~arcsec$^{-2}$) galaxies.
 The vast majority of their galaxies have 0.4 $< g$-$r<$1.0~mag in color ( as shown in Figure 5 in Bell03 paper).
 RC15 MLCR is based on a representative sample of nearby bright galaxies 
 with apparent $B$-band magnitudes $>$16~mag from the Virgo cluster.
 
 IP13 and Z09 are purely theoretical MLCRs.
 IP13 MLCR is based on a combination of simple stellar populations (SSPs) from the isochrone data set of the Padova group
 which includes a revised (new) prescription for the thermally pulsing asymptotic giant branch (TP-AGB),
composite stellar populations (CSP) which are generated by convolving SSPs 
 according to exponentially declining (or increasing) star formation histories (SFHs),
 and disc galaxy models from \citet{Portinari2004}.
 Z09 MLCR is based on a Monte Carlo
library of 50 000 SSPs from the 2007 version of \citet{Bruzual2003} (CB07),
 assuming a two-component SFH of a continuous, exponentially declining mode with random bursts superimposed 
 and also includes a revised (new) prescription for TP-AGB phase.
 Although they strive to construct a representative sample of
 the whole galaxy populations, they are all biased to HSB and redder galaxies to varying degrees. 
 For example, B03 sample lacks for sufficient LSBGs and bluer galaxies ($g$-$r <$0.4~mag),
which would cause a weak constraint for the MLCR in the bluer color part,
but be more constrained by the redder galaxies with $g -r >$0.4~mag.
However, our MLCR is calibrated for a sample of LSBGs (most of which have $\mu_{r} >$21~mag~arcsec$^{-2}$ and $r >$17.5~mag),
and 73$\%$ of the sample is bluer than $g$-$r$=0.4~mag with the rest between 0.4 $< g$-$r<$ 1.0~mag.
This work is hitherto the first attempt to test the MLCR on LSBGs
which are proposed to be potentially different from HSB galaxies.
 In this case, we additionally overplot the H16 MLCR (as magenta dashed lines) 
 which is based on a sample of 34 dwarf irregular (dIrr) galaxies
 in the corresponding panels in left column of Figures ~\ref{fig:m2l_color_gr},
 since H16 only gives the MLCR of log $\gamma_{*}^{j}$ ($j$=$g$ and $r$) with $g$ - $r$ color.
It seems that the H16 MLCR is even much flatter than the B03 MLCR.
Therefore, the distinction/disparity between sample properties may lead to 
subtle/significant differences between the MLCRs.

In addition, both our and B03 MLCRs are based on photometry 
measured on SDSS images, but the methods of photometry between
us is distinguishing.
B03 adopts the SDSS Petrosian (SDSS Petro) magnitudes,
measured within an circular aperture that is twice the radius
at which the local surface brightness is 1/5 of the mean 
surface brightness within that radius.
As we mentioned in $\S$ ~\ref{sec:phot},
the SDSS Petro has been reported to underestimate the magnitudes of bright galaxies 
by $\sim$0.2~mag \citep{He2013} and those of dwarf galaxies by up to $\sim$0.5~mag \citep{Lisker2007}
due to its sky estimate algorithm that tends to subtract the LSB parts of galaxies
as part of sky background.
To correct this problem, B03 roughly subtracts 0.1~mag from the SDSS Petro magnitudes 
for their sample galaxies.
However, for our LSBGs, 
we take a more accurate sky estimate method \citet{Zheng1999,Wu2002,Du2015}
in advance to generate an 'unbiased' 2-D sky background for each of the SDSS images. 
The sky-subtracted images are then fed into the SExtractor code to measure
the magnitudes of our LSBGs within the Kron elliptical aperture (SEx AUTO; recommended by SExtractor),
which is distinct from the SDSS Petrosian circular aperture that B03 adopts. 
\citep{Hill2011} tested the mean (median) difference between SDSS Petrosian circular aperture magnitude 
and the SExtractor Kron elliptical aperture magnitude for a same galaxy sample
and found the offset is 0.04(-0.01), 0.03(0.01), 0.02(0.01), 0.04(0.01), and 0.02(-0.01)~mag
in $u$, $g$, $r$, $i$, and $z$ bands, respectively.
Obviously, the offset of magnitudes arising from difference of the two aperture definitions are small.
Here we also tested the difference between SEx AUTO and SDSS Petro (from SDSS DR7 photometric catalogue)
magnitudes for our sample. As shown in Figure ~\ref{fig:mag_SDSSour},
the mean(median) offset of SEx AUTO from SDSS Petro magnitudes
is 0.18(0.26), 0.16(0.23), 0.09(0.14), 0.31(0.37), 0.52(0.62)~mag,
respectively, in $u$, $g$, $r$, $i$, and $z$ bands.
Compared to the results of \citet{Hill2011},
these offsets for our LSBGs are larger, which is mainly due to our correction for
the underestimation of SDSS Petro magnitude by using a different but better sky-subtraction recipe ($\S$ ~\ref{sec:phot})
from the SDSS one. Apparently, these offsets in magnitudes for our sample 
are reasonably consistent with the expected correction value for sky subtraction provided
in previous literature \citep[e.g.][]{He2013,Lisker2007},
but it would also cause the offset of zero-point (color and log $\gamma_{*}$) between B03 and our MLCRs
because the B03 MLCR subtracts 0.1 dex from the SDSS Petro magnitudes of their galaxies to correct the sky subtraction problem of the SDSS photometry. 
However,   for our sample, our mean correction value for SDSS Petro magnitudes in $g$, $r$, $i$, and $z$ bands
are -0.16, -0.09, -0.31, and -0.52~mag, respectively, while according to B03 method, the correction value would be
-0.1~mag for all these 4 bands. 
So our correction would result in a mean offset of $\sim$ -0.07 in $g$ - $r$ and 0.15~mag
in $g$ - $i$ color zeropoints from B03 correction for our sample, 
which implies that our sample would systematically shift redwards by $\Delta$($g$-$r$)$\sim$0.07~mag in all panels of 
the left colomn in Figure ~\ref{fig:m2l_color_gr},
but shift bluewards by $\Delta$($g$-$r$)$\sim$0.15 mag in all panels of the left column in Figure ~\ref{fig:m2l_color_gi},
if the B03 correction was applied to our sample.
In this similar way, we can derive 
that our sample would shift upwards by $\Delta$log$\gamma_{*}^{j} \sim$ 0.02, 0.0, 0.08, 0.17 dex for $j$=$g$,$r$,$i$, and $z$ 
in the corresponding panels in Figures ~\ref{fig:m2l_color_gr} and ~\ref{fig:m2l_color_gi},
if the B03 correction was applied to our sample.

So in most panels, the systematic shifts of our sample blueward in color and especially upward in stellar mass-to-light ratio
would reduce the offset between our sample and the B03 MLCR, although they can not completely eliminate
the disparity between our sample and the B03 MLCR obviously shown in Figures ~\ref{fig:m2l_color_gr} and ~\ref{fig:m2l_color_gi}.


 

\subsection{Fitting techniques}\label{sec:fitting_method}
The MLCRs are given by fitting the galaxy data to linear lines.
In $\S$ ~\ref{sec:ms}, we fit linear MLCRs 
(black solid lines in the left column of both Figures ~\ref{fig:m2l_color_gr} and ~\ref{fig:m2l_color_gi}) 
to our LSBG data by using a bi-square weight technique (biweight), 
which uses the distance perpendicular to the bisecting line of the data to calculate weights
of data points for fitting and was also adopted by B03 MLCRs.
Since diverse fitting techniques may result in distinguishing coefficients for the fitting line,
we show the fitting lines by using another two different line fitting techniques for our LSBG data
in the right column of Figures ~\ref{fig:m2l_color_gr} and ~\ref{fig:m2l_color_gi}.
The dark green line in each panel is given by the MPFITEXY, an IDL fitting procedure, which finds the best-fitting straight line 
through data with errors in both coordinates taken into consideration of calculating the weights for 
each data point.
The orange line in each panel is given by a direct fitting method, 
which directly fits the LSBG data to a linear model by minimizing
the $\chi^{2}$ with no weights considered for fitting.
For a clear comparison, the black solid line previously given by `biweight' fitting technique
is also overplotted in each panel of the right column of 
Figures ~\ref{fig:m2l_color_gr} and ~\ref{fig:m2l_color_gi},
and the coefficients of the MLCRs given by the three different line fitting methods for our LSBG data 
are also tabulated in Table ~\ref{tab:m2l_color_methods}.
Obviously, in each panel, the `mpfitexy' (dark green line) and 'direct' (orange line)
techniques give closely consistent fitting lines, 
which are much flatter than the `biweight' fitting line (black solid line),
agreeing much better with the flatter B03 MLCR in slope iin
each panel in the right column of Figures ~\ref{fig:m2l_color_gr} and ~\ref{fig:m2l_color_gi}.
This reveals that the results of the fitting line are more or less dependent on the fitting technique,
and we could not guarantee that 
our fitting technique is completely the same as those techniques used by other literature MLCRs.
  
\begin{table*}
\caption{The coefficients for the MLCRs from three different line fitting methods for our LSBG data in the form of log($\gamma^{*}$)=$a_{\lambda}$+($b_{\lambda}\times$color).}
\label{tab:m2l_color_methods}
\begin{center}
\begin{tabular}{lcccccccc}
\hline
\hline
color & $a_{g}$ & $b_{g}$  &$a_{r}$ & $b_{r}$ &$a_{i}$ & $b_{i}$ &$a_{z}$ & $b_{z}$\\
\hline
\hline
\multicolumn{9}{l}{\bf{biweighted (black solid line)}}\\
$g-r$&-0.857&1.558&-0.700&1.252&-0.756&1.226&-0.731&1.128\\
$g-i$&-1.152&1.328&-0.947&1.088&-0.993&1.057&-0.967&0.996\\
 \hline
 \multicolumn{9}{l}{\bf{direct (orange solid line)}}\\
 $g-r$&-0.709&1.072&-0.526&0.672&-0.607&0.732&-0.537&0.490\\
$g-i$&-0.924&0.932&-0.719&0.691&-0.691&0.530&-0.741&0.606\\
\hline
\multicolumn{9}{l}{\bf{mpfitexy (dark green solid line)}}\\
$g-r$&-0.743&1.141&-0.556&0.732&-0.635&0.775&-0.550&0.522\\
$g-i$&-0.964&0.976&-0.744&0.708&-0.723&0.563&-0.762&0.633\\                                                                  
  \hline
  \end{tabular}
   \end{center} 
 \end{table*}
\subsection{SED model ingredients}
The basis of using SED-fitting method to estimate stellar mass or stellar mass-to-light ratio is
that galaxies can be viewed as a convolution of simple stellar populations (SSPs)
of different ages and metalicities,
according to a specific star formation and chemical evolution history (SFH). 
So far there are various stellar population (SP) models,
e.g., \citet[hereafter Vazdekis model]{Vazdekis1996},
\citet[hereafter PEGASE model]{Fioc1997}, 
\citet[hereafter BC03 model]{Bruzual2003},
the 2007 version of BC03 (hereafter CB07),
and \citet[hereafter FSPS model]{Conroy2009}.
These models are based on various stellar evolutionary tracks/or isochrones,
stellar libraries, the prescriptions for the late stage of evolution,
such as Thermally-Pulsing Asymptotic Giant Branch (TP-AGB) phase,
assuming various initial mass functions (IMFs) and SFHs.

For the MLCRs in the figures, 
RC15 uses two independent SP models of BC03 (RC15(BC03))
and FSPS (RC15(FSPS)), as shown in the MLCR figures, 
FSPS model (grey dashed line) always gives
flatter MLCR than the BC03 model (cyan dashed line) in each panel. 
This demonstrates that the choice of SP model matters a lot,
which would result in distinguishing MLCRs, especially in slope.
B03 adopts PEGASE model, incorporating an `old' prescription \citep{Girardi2000, Girardi2002} for TP-AGB stars. 
IP13 used the new Padova model (SSP model generated from the isochrones of Padova group),
and Z09, H16, and our MLCRs are all based on the CB07 model, 
which all incorporates a `new' prescription \citep{Marigo2007,Marigo2008} for TP-AGB stars.
The 'new' treatment for TP-AGB phase 
includes a larger contribution from TP-AGB stars
that change little to the optical luminosity dominated by main-sequence stars,
but greatly enhances the predicted redder/near-infrared luminosity of galaxies, 
and finally leads to lower $\gamma^{*}$ in redder/near-infrared bands than
bluer bands. This is why the disparity between B03 (red dashed line) and our MLCRs (black solid line) 
in the MLCR figures
is becoming larger in panels of red bands of $i$ and $z$ bands than those of $g$ and $r$ bands
for our sample.

IMF is critical for determining the $\gamma^{*}$ of a galaxy from SED-fitting method.
The extant IMFs are mostly determined for the Milky Way in the Solar neighborhood,
while the IMF of external galaxies is in principle unknown \citep{Courteau2014}.
The various IMFs, e.g., the Salpeter IMF \citep{Salpeter1955}, 
Kroupa IMF \citep{Kroupa1998, Kroupa2001}, and Chabrier IMF \citep{Chabrier2003},
 differ mainly in the slope for low mass stars,
 which serve mostly to change the mass without much changing the luminosity or color.
So an IMF that includes more low mass stars, such as the Salpeter IMF,
yields a higher $\gamma^{*}$ at a given color, 
since the large number of low-mass stars would greatly increase the stellar mass, 
but hardly affect the luminosities or colors of galaxies as these stars are too faint.
Hence, different IMFs would result in offsets in zero-point of the MLCRs,
even though the relation slopes may remain unchanged \citep{Bell2001, Bell2003, Courteau2014}.
In our modeling, we assumed a Chabrier IMF which gives small number of low-mass stars. 
However, the B03 MLCR assumed a ``diet ''  \citet{Salpeter1955} IMF, 
which gives a larger number of stars in the low mass end of IMF
and thus yields a higher $\gamma^{*}$ at a given color
than the Chabrier IMF. 
The IP13 MLCR uses a \citet{Kroupa1998} IMF.
Although we have converted both the B03 and IP13 MLCRs
from its original IMFs to a Chabrier IMF by adding a correction value to the originally predicited log $\gamma_{*}f$ 
for comparison in the left column of the MLCR figures (Figures ~\ref{fig:m2l_color_gr} and ~\ref{fig:m2l_color_gi}).
If we choose the correction value to be -0.093 dex for B03 MLCR following \citet{Gallazzi2008,Zibetti2009} 
and 0.057 dex for IP13 following \citet{Herrmann2016}, the IMF-corrected B03 MLCR (red dashed line) shown in 
the MLCR figures is still higher than our MLCR;
however, the IMF-corrected IP13 (green dashed line) appears more consistent with our MLCR in zeropoint
in each panel of the MLCR figures.
If the correction values changed, the zero-points of both MLCRs
would shift accordingly.
So the zero-points of the IMF-converted B03 and IP13 MLCRs in the figures should be still
dependent on the uncertainties of the correction values which we do not know for sure.
Besides, we are not certain about whether the correction values used here are accurate,
as -0.093 dex is used in H16 while -0.15 is used in RC15.
Similar to our MLCR, the Z09 (blue dashed line) and RC15 (BC03) (cyan dashed line) MLCRs
also use a Chabrier IMF, so they are obvious to be more consistent with our MLCR in zero-point
than other literature MLCRs in the figures.
Unlike the RC15(BC03), the RC15(FSPS) MLCR (grey dashed line) has an relatively larger offset
from our MLCR, because it uses a different stellar population model of FSPS, although it also uses a Chabrier IMF.
This will be discussed later.
Although H16 (magenta dashed lines in the first two panels) MLCR uses a Chabrier IMF,
it is much higher (like B03) than our MLCR in each panel. This should be due to the different
line fitting technique (as discussed in \S~\ref{sec:fitting_method}) and other different SPS model ingredients
which will be discussed later.

The SFH regulates the star formation and chemical enrichment with time.
The choice of the SFH, in particular whether it is rising, declining, or bursty, 
can significantly change the best-fit stellar mass, 
by perhaps as much as 0.6 dex in extreme cases 
\citep{Pforr2012,Conroy2013}.
IP13 considers an exponentially declining (or increasing) SFH
in the form of $\rm \Psi$($t$)=e$^{-t/\tau}$.
The declining ones are modelled with e-folding time-scales $\tau$ 
ranging from 1.55 Gyr to $\infty$ (constant star formation rate, SFR)
and the increasing ones are modelled with negative values of $\tau$
ranging from -50.00 to -1.00.
H16 MLCR is based on a library of multi-component complex SFHs 
which is created for dwarf irregular galaxies by \citet{Zhang2012}
and totally different from the commonly used 2-component SFH model
from literature.
Our MLCRs, RC15 MLCRs and Z09 MLCRs are all based on the 2-component SFH model 
(an exponentially declining SFHs 
with random bursts superimposed), which incorporates a variety of bursty events and allows young ages (of a few Gyr).
Whereas B03 considers relatively smooth SFHs starting from 12 Gyr in the past, 
and limits the strength of star bursty events (which are simultaneously constrained to only happen
in the last 2 Gyr) to $\leqslant$ 10 percent by stellar mass.
As a recent burst of star formation will dramatically
lower the mass-to-light ratio of a total stellar system than a smooth star formation model
by up to 0.5 dex \citet{Courteau2014},
omitting or less burst components
from SFH models will bias B03 MLCRs-based stellar mass-to-light ratio to higher values\citep{Roediger2015}.

%

\subsection{Effect of surface brightness on fitting relations}

Since a lower surface brightness is the characteristic 
that distinguishes LSBGs from HSBGs (mostly normal spiral galaxies),
we here check whether the MLCRs
have dependence on galaxy surface brightness,
besides the external factors discussed above.
We divide our LSBG data into three central surface brightness bins 
($\mu_{0,B}\leqslant$23.5, 23.5 $<\mu_{0,B}<$24.5, and $\mu_{0,B}\geqslant$24.5).
These $B$-band central surface brightness are all measured in \citet{Du2015}. 
Then, we fit the MLCR for galaxies in each bin using the `bi-weight' line fitting technique.
The MLCR for each bin is over-plotted (as three brown lines of different styles) in each panel
in the right column of Figures ~\ref{fig:m2l_color_gr} and ~\ref{fig:m2l_color_gi}.  
It is obvious that the MLCRs slightly flatten with the lower central surface brightness,
as shown in Figure ~\ref{fig:slope_mu}. 
However, the change of MLCR due to the change of central surface brightness
(represented by the differences between black and brown lines) is small,
which is far less than that due to the change of fitting techniques
(shown by the larger offset between the black and dark green/orange lines)
in the right column of the MLCR figures. 

In this case, we think the minor differences between our MLCR (for LSBGs)
and the literature MLCRs (for the whole galaxy population, HSBGs, or dwarfs)
are more caused by the combination of differences in SED fitting models (IMF, SFH, and SPS),
photometric zero-point of data, and line fitting techniques, rather than the central surface brightness
of galaxies themselves.
As shown in the left column of the MLCR figures, 
our MLCR (black solid line) for LSBGs is among the literature MLCRs
in each panel, so it is expected to be generally consistent with those literature MLCRs for various samples,
if those main factors of difference in photometric methods, line fitting methods,
and models (IMF, SFH, and SPS) are taken into account.
This could be further in evidence by the best consistency
between our MLCR and Z09 MLCR (blue dashed line),
especially in Figure ~\ref{fig:m2l_color_gi},
which is based on the same IMF, SPS model (CB07) and SFH as our MLCR
except for based on a theoretical library of galaxy samples.

\section{Comparison with Huang12's stellar mass for dwarf galaxies}
In this section, we hope to assess our MLCRs by comparing the predicted stellar masses
with those estimated in an independent paper \citet[Huang12]{Huang2012} for dwarf galaxies
with comparable properties.

Huang12 defined a sample of gas-rich dwarf galaxies,
which have log $M_{H{\sc{i}}}<$7.7 and $W$50 $<$ 80 km s$^{-1}$
from the $\alpha$.40 HI survey catalogue.
They give stellar masses for this sample galaxies 
by fitting their SEDs consisting of $GALEX$ (FUV NUV) and SDSS ($ugriz$) photometric bands 
to the BC03 model, assuming a \citet{Chabrier2003} IMF
and a continuous SFH with random bursts superimposed, 
which are the same as our assumptions.
Besides the SED fitting results, Huang12 also predicted the stellar masses
according to the B03 MLCR ($i$ v.s. $g$-$r$). 
In comparison, their SED fitting yields a median log $M_{*}/M_{\odot}$=7.45
while the B03 MLCR (converted to Chabrier IMF) 
gives a considerably higher median of 7.73,
which overestimates the stellar mass by $\sim$0.28 dex.Using Huang12's criteria, 
we derive a sample of dwarf galaxies out of our LSBGs.  
Our MLCR ($i$ v.s. $g$ - $r$) yields a median log $M_{*}/M_{\odot}$=7.36
while B03 MLCR (converted to Chabrier IMF) 
gives a considerably higher median of 7.69 for our dwarf galaxies,
which overestimates the stellar mass by $\sim$0.33.
This result is very similar to Huang12's result
in terms of both the stellar mass value and the offset value from B03 predictions,
which gives confidence for using our MLCR, especially for dwarf galaxies.
The offset should be mainly caused by the ingredients discussed in $\S$~\ref{sec:discuss},
and Huang12 claimed that it is mainly due to the different 
SFH adopted by Bell03, which does not fully account for the impact
of bursty behavior in dwarf galaxies.

\section{Summary and Conclusion}
We obtained stellar masses, $M_{*}$, and stellar mass-to-light ratio, $\gamma_{*}$,
for a sample of low surface brightness galaxies (LSBGs) 
by fitting their spectral energy distributions (SEDs) covering eleven ultraviolet, optical and near-infrared 
photometric bands to the stellar population synthesis (SPS) model using the MAGPHYS code \citet{da Cunha2008}. 
The derived $M_{*}$ for this sample spans from log$M_{*}/M_{\odot}$= 7.1 to 11.1 dex,
with a mean log $M_{*}/M_{\odot}$ = 8.47 and a median of 8.48 dex,
showing that these LSBGs have a systematically lower $M_{*}$ than normal galaxies.
The $\gamma^{*}$ for this sample slightly decreases
from $r$-band to redder wavelength bands for our LSBG sample, similar to 
the declining trend of $\gamma^{*}$ from short to longer wavelength
for normal star-forming galaxies,
and the $\gamma^{*}$s vary little with absolute magnitude,
but slightly increase with higher $M_{*}$ for our LSBG sample.
This increasing trend is stronger in bluer bands, 
with a steepest slope in $u$ but nearly flat slope in the $K$ band.

We then fitted the stellar-mass-to-light - color relation (MLCR) for the LSBG sample. 
The log $\gamma_{*}^{j}$ ($j$=$g$, $r$, $i$, and $z$) have the relatively
tightest relations with the optical colors of $g-r$ and $g-i$ for our LSBG data. 
Compared with the literature MLCRs, our MLCRs are consistently
among those literature MLCRs
that are converted to the same IMF.
The minor differences could be more due to the differences
in SED models (IMF, SFH, SPS model), photometric zeropoints,
and line fitting techniques, but depend little on the galaxy surface brightness.
This may give a possible implication that most of our LSBGs might share the generally 
similar properties in star formation and evolution with the normal galaxies.


\acknowledgements

We appreciate the anonymous referee for his/her constructive comments,
which makes this paper strengthened. 
DW is supported by the National Natural Science Foundation of China (NSFC)
grant Nos. U1931109, 11733006, the National Key R$\&$D Program of China 
grant No. 2017YFA0402704, and the Young Researcher Grant funded by National 
Astronomical Observatories, Chinese Academy of Sciences (NAOC).
CC is supported by the NSFC grant No.11803044, the Young Researcher Grant funded by NAOC,
and also supported in part by the Chinese Academy of Sciences (CAS) 
 through a grant to the CAS South America Center for Astronomy (CASSACA) in Santiago, Chile.
ZZ is supported by the NSFC grant No. 11703036.
WH is supported by the National Key R$\&$D Program of China 
grant No. 2017YFA0402704, and the NSFC grant Nos.11733006. 
 



%

\vspace{5mm}
\facilities{GALEX, SDSS, UKIDSS}


\software{SExtractor \citep{Bertin1996}, 
MAGPHYS \citep{da Cunha2008}
          }

\clearpage

\begin{figure}[ht!]
\centering
\includegraphics[width=0.8\textwidth]{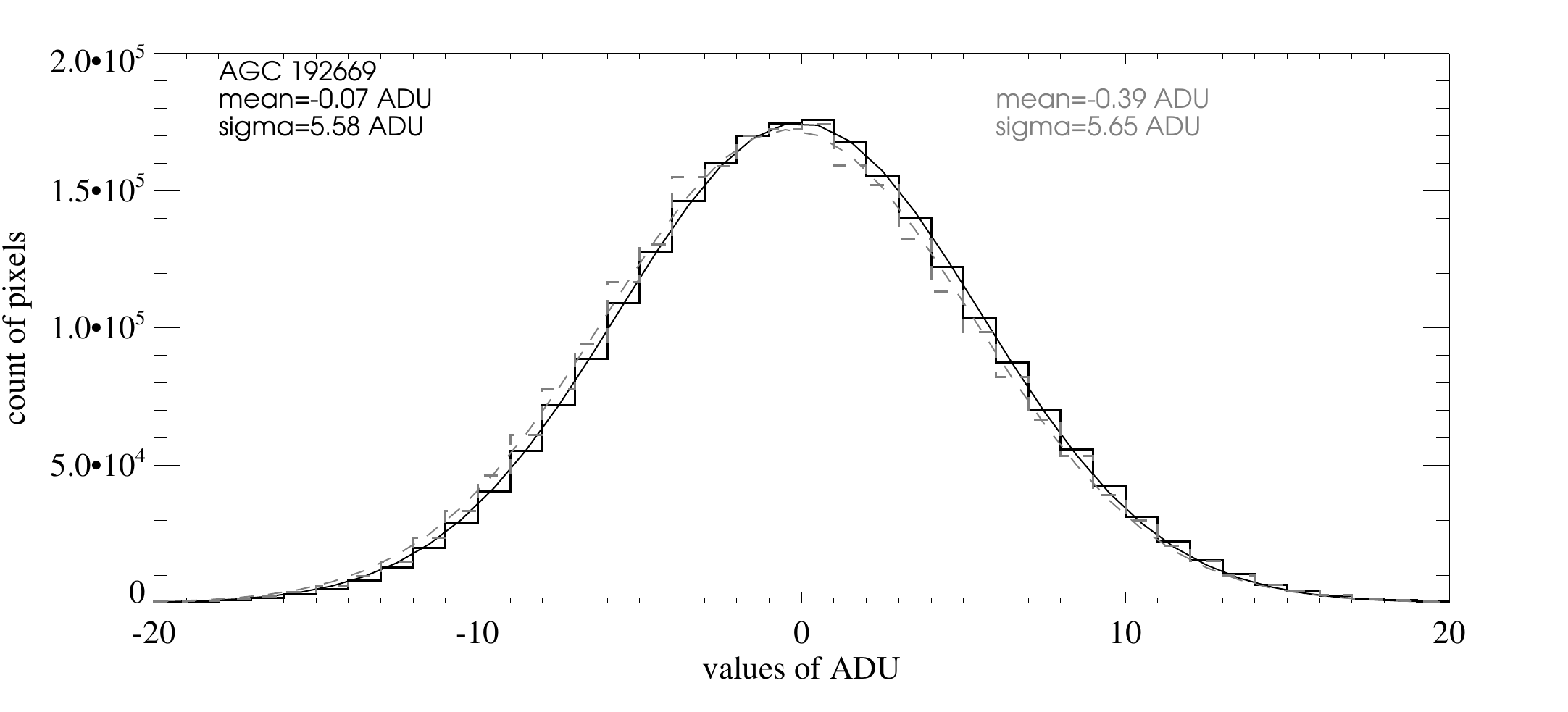}
\caption{Distributions of the pixel values of all the unmasked region (with object masked during our sky estimation process)
of the sky-subtracted image frame of AGC 192669.
The solid black line represents sky-subtracted frame by our sky subtraction method of piecewise row-by-row
and column-by-column fitting, and, for comparison, the dashed grey line are
for the frame without our accurate sky background subtraction but only with the simple
mean value of all the unmasked region subtracted (similar to the SDSS method which gives a single sky value for an image frame). 
It demonstrates that the mean background value is more close to zero after applying our sky subtraction method.}\label{fig:sky_subtraction}
\end{figure}

\begin{figure}
\centering
\includegraphics[width=1.0\textwidth]{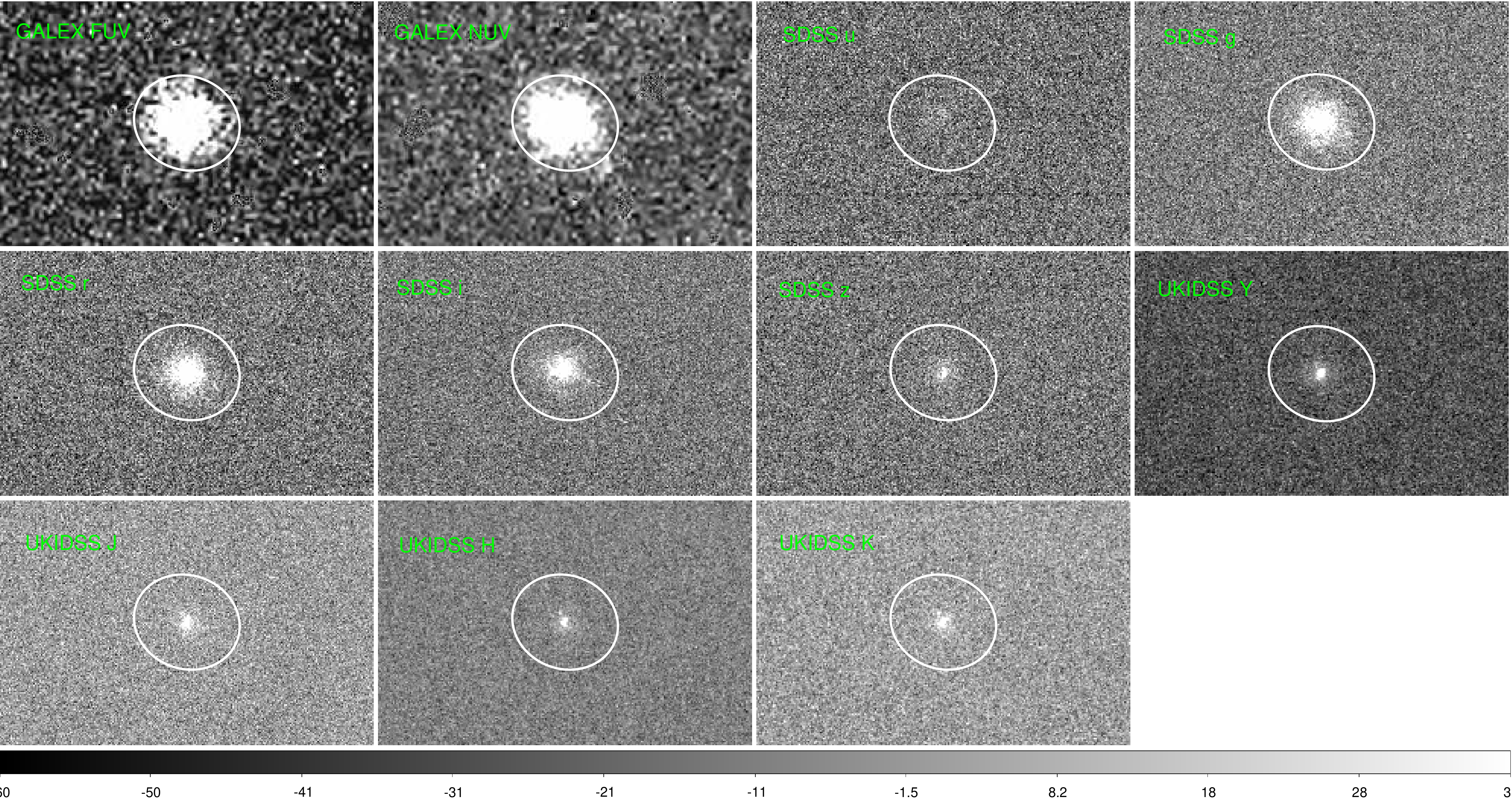}
\caption{The homogeneous photometric apertures of an example LSBG AGC 192669
in multi-bands from GALEX $FUV$, $NUV$, SDSS $u$, $g$, $r$, $i$, $z$, UKIDSS $Y$, $J$, $H$ and $K$.}\label{fig:phot}
\end{figure}

\begin{figure}
\centering
\includegraphics[width=0.8\textwidth]{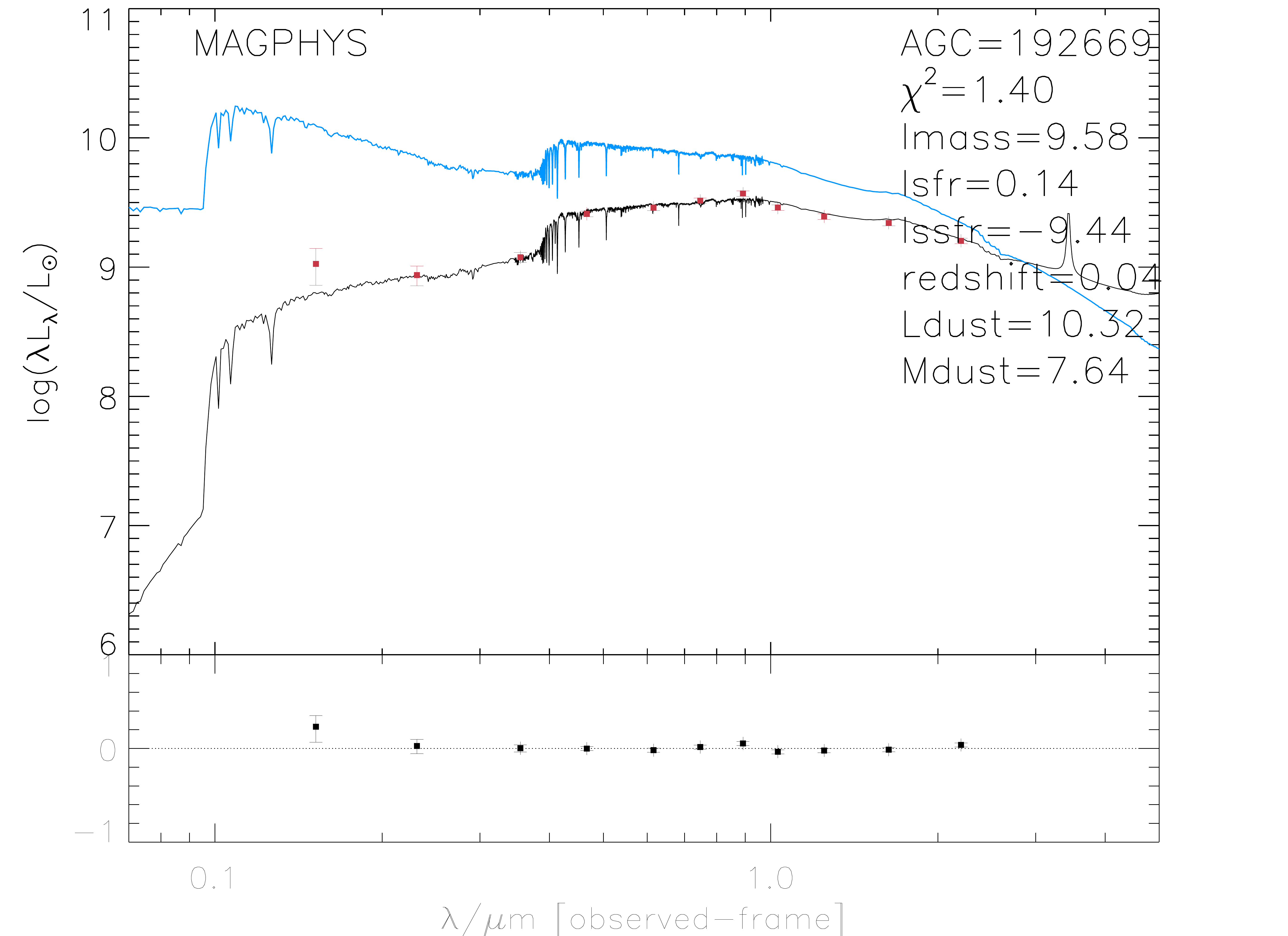}
\caption{The best fit for the SED of an LSBG galaxy AGC 192669 by using the MAGPHYS (top). 
The best-fitting model (black curves) to the observed SED (red filled squares) is shown in their upper panel, 
and the residual between model and observation are shown in the lower panels. 
The blue curve in the MAGPHYS panel shows the unattenuated model spectrum.  
Additionally, the stellar mass ($lmass$), star formation rate($lsfr$), log specific star formation rate ($lssfr$),
redshift, and dust ($Ldust$ and $Mdust$) given by the the best-fitting model are listed in the panel.}
 \label{fig:sed_fitting}
\end{figure}

\begin{figure}
\centering
\includegraphics[width=0.85\textwidth]{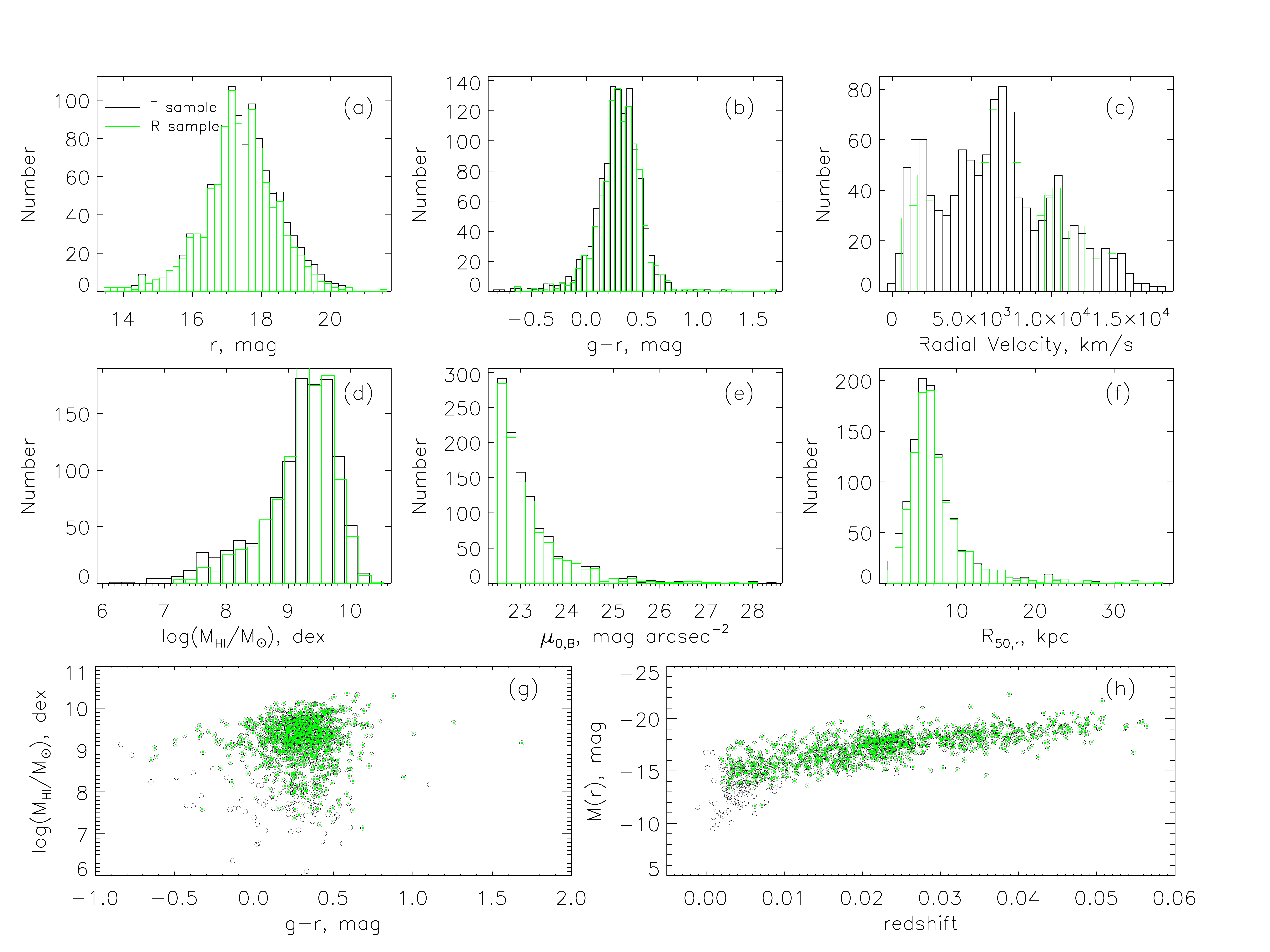}
\caption{Properties of the R sample (green) in comparison with the T sample (black)
in terms of (a) $r$-band magnitude, $r$, (b) optical color, g-r, (c) radial velocity, in km/s,
(d)H{\sc{i}} mass, log(M$_{H{\sc{i}}}$/M$_{\odot}$), in dex (e) central surface brightness in $B$ band, $\mu_{0,B}$,
in mag~ arcsec$^{-2}$,
(f)effective radius, $R_{50,r}$, in kpc, (g)H{\sc{i}} mass versus g-r, and
(h)$r$-band absolute magnitude versus redshift.}\label{fig:subsample}
\end{figure}

\begin{figure}
\centering
\includegraphics[width=0.85\textwidth]{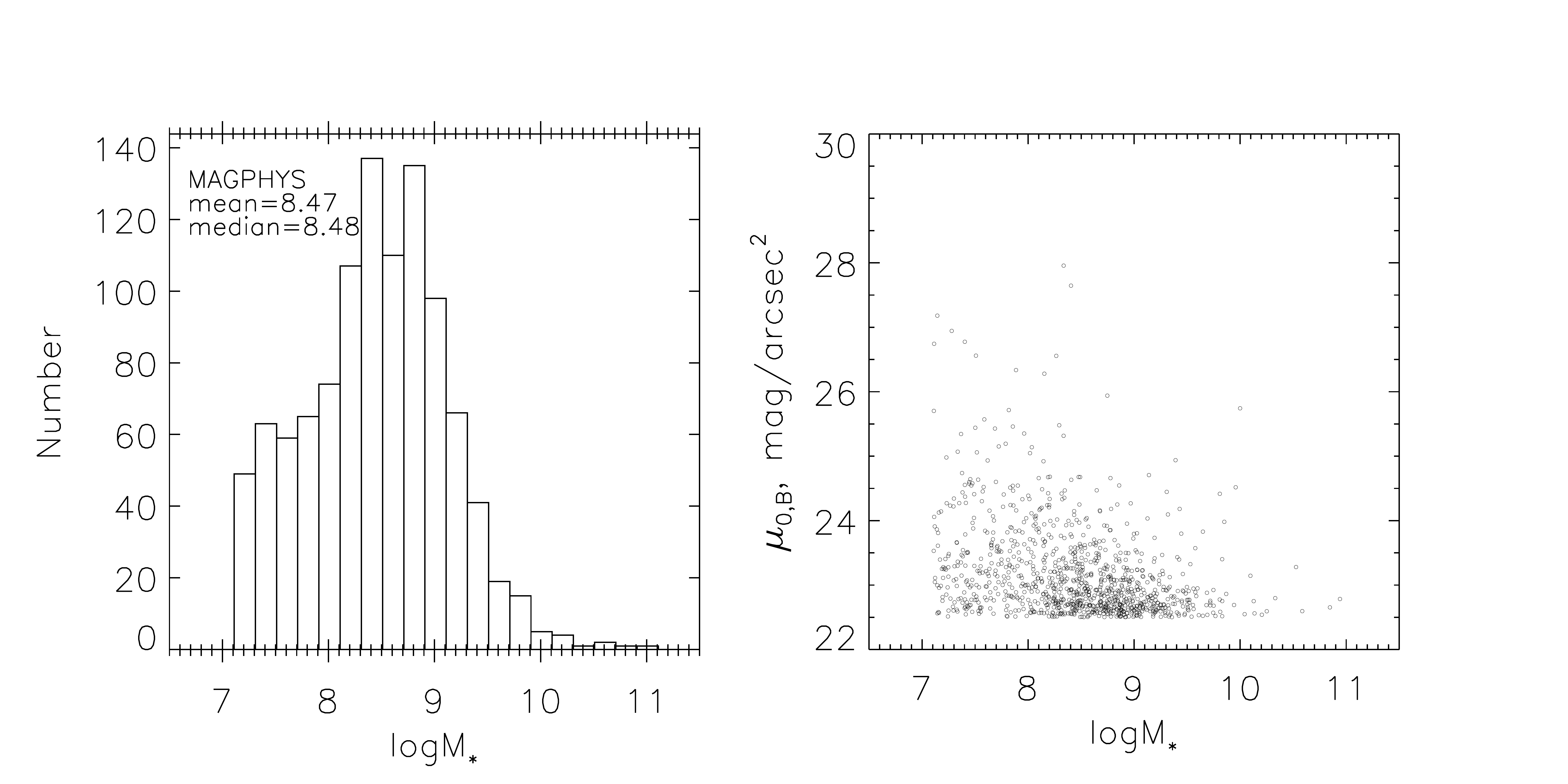}
\caption{Stellar masses for the R sample derived from the SED fitting using MAGPHYS. 
The left panel is the distribution histogram of stellar mass,
 and the right panel shows the stellar mass, log$M_{*}$, versus central surface brightness, $\mu_{0,B}$,
 that is measured in \citet{Du2015} and given in \citet{Du2019}.}
 \label{fig:mstar_distri}
\end{figure}

\begin{figure}
\centering
\includegraphics[width=1.0\textwidth]{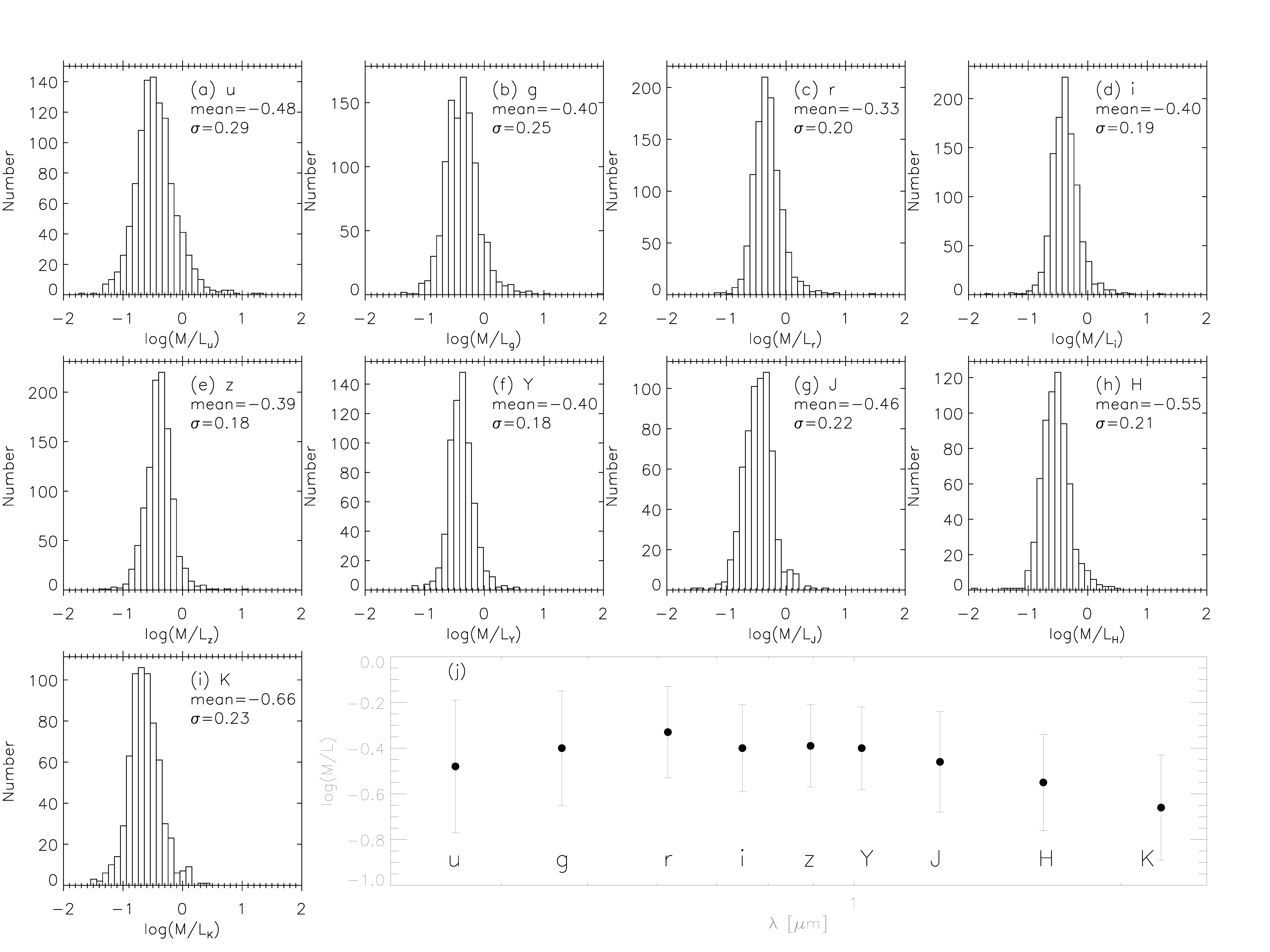}
\caption{The distribution of stellar M/L ratios based on luminosities in each band from
$FUV$ to $K$ bands.}
\label{fig:m2l_distr}
\end{figure}

\begin{figure}
\centering
\includegraphics[width=1.0\textwidth]{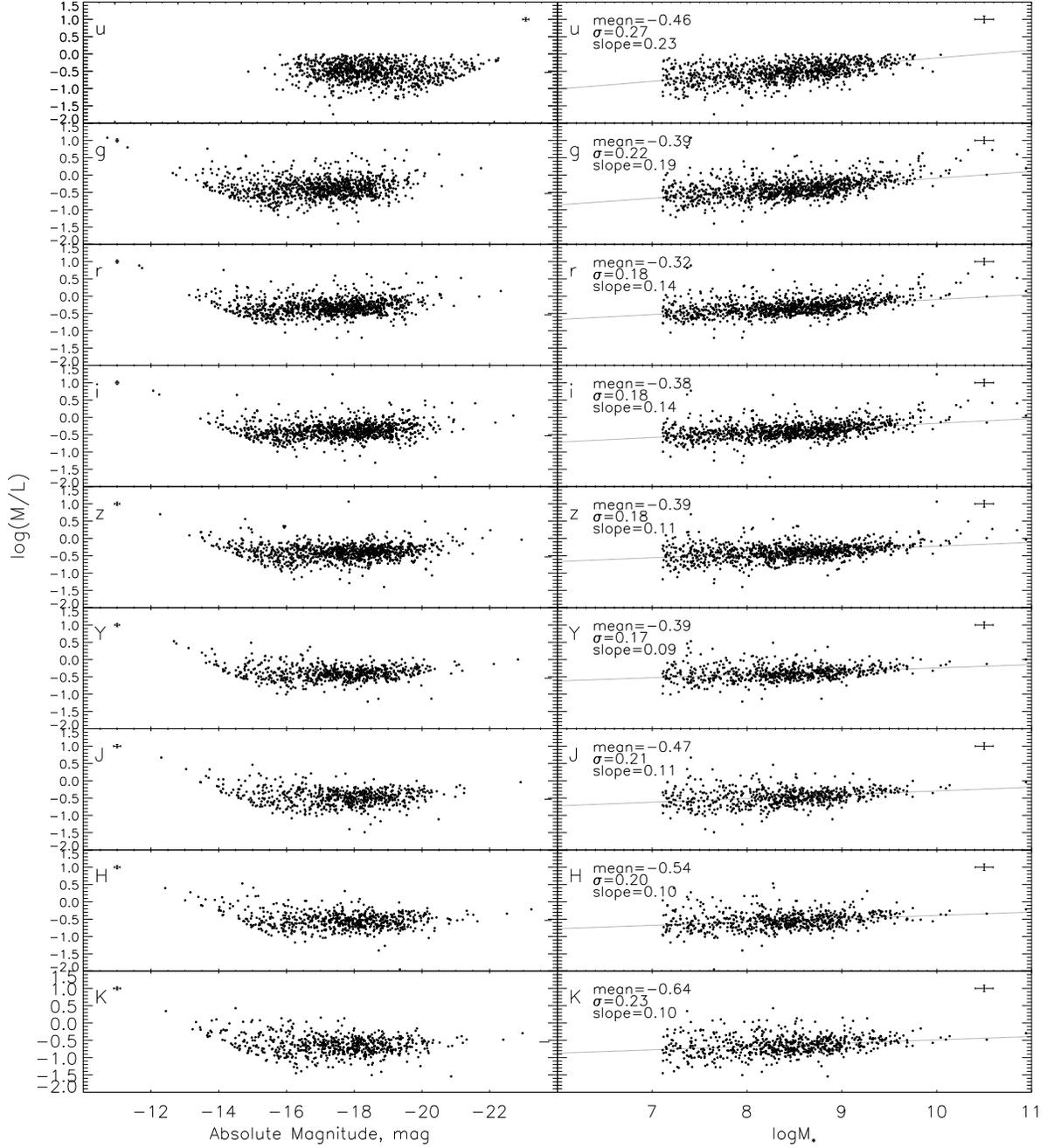}
\caption{The R LSBG sample in the $\gamma^{*}$ -- absolute magnitude (left) and 
$\gamma^{*}$ -- stellar mass (right) planes. 
Here, stellar masses measured
in $ugrizYJHK$ bands are shown from top to bottom, respectively. 
To prevent the picture from a vague blob, 
only the average error bars in both axises are given in each panel. 
It should be noted that the error of absolute magnitude in the $x$-axis
is from a mathematic propagation of the errors of galaxy magnitude (measured by SExtractor) and
galaxy distance (directly given in ALAFLFA catalogue \citep{Haynes2018}), 
and the error of $\gamma^{*}$ in the $y$-axis is from a mathematic propagation of the errors of stellar mass, M$_{*}$,
(estimated by the MAGPHYS code) and galaxy luminosity (measured by SExtractor).} \label{fig:m2l}
\end{figure}

\begin{figure}
\centering
\includegraphics[width=1.0\textwidth]{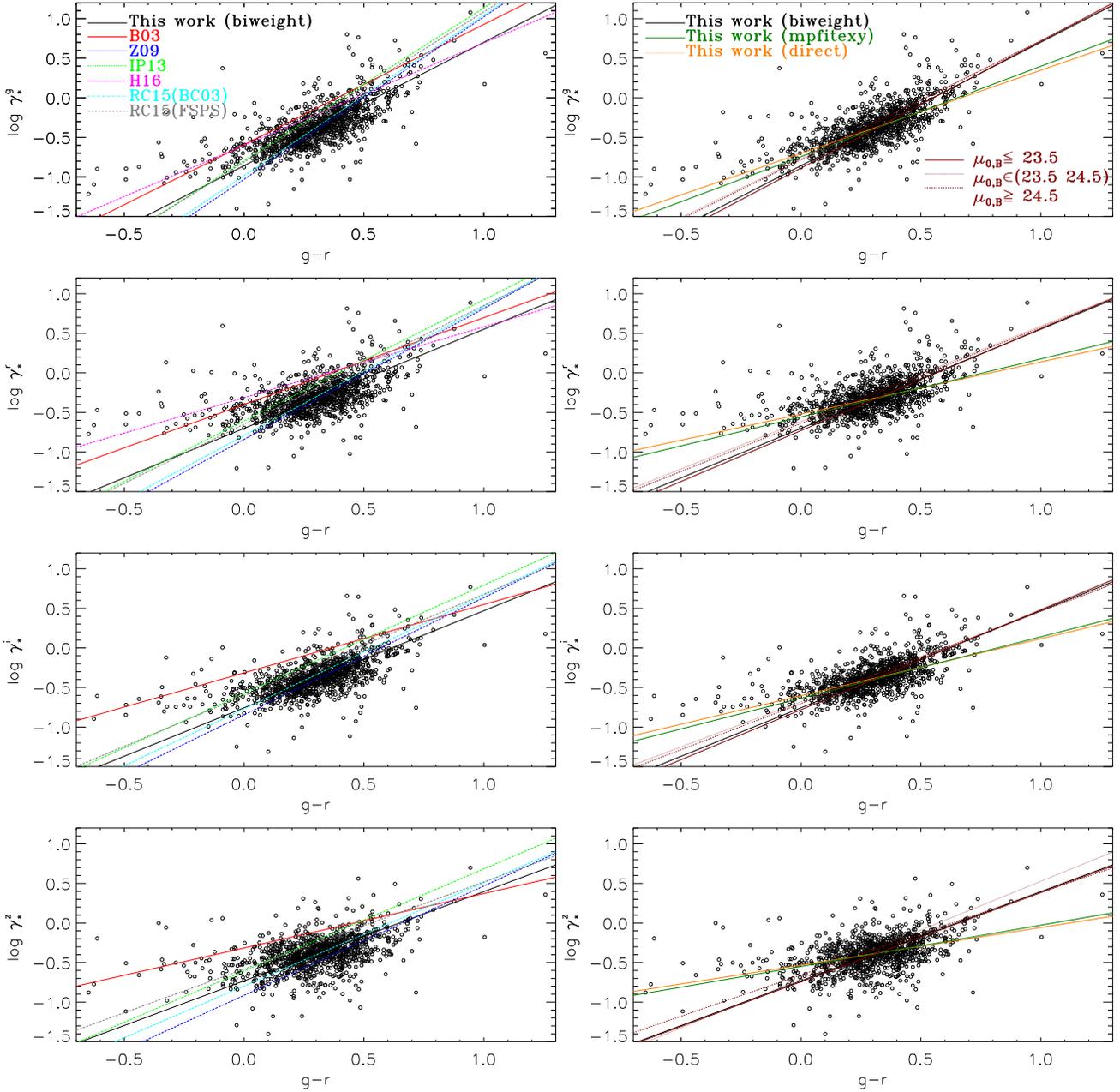}
\caption{Comparison of estimated $g$-,$r$-,$i$- and $z$-band stellar $\gamma^{*}$ ratios
as a function of $g-r$ color for LSBGs in this paper (black circles). In all panels, we show our correlation
by a ``robust" bi-square weighted line fit solid black line) with the assumption of
a \citet{Chabrier2003} IMF and the CB07 stellar population model.
For comparison, we overplot the correlations of \citet[B03]{Bell2003} as the red line, \citet[Z09]{Zibetti2009} as the blue line, \citet[IP13]{Into2013} as the green line, 
\citet[H16]{Herrmann2016} as the magenta line, \citet{Roediger2015} based on the BC03 model (RC15(BC03)) as the cyan line, and \citet{Roediger2015}
based on the FSPS model (RC15(FSPS)) as the grey line in each panel. All the CMLRs from literature
shown here are under or have been corrected under the \citet{Chabrier2003} IMF.
We note that the color - M/L correlations are in the form of log$_{10} (M/L)$ = $a_{\lambda}$ + $b_{\lambda}\times$color.}
\label{fig:m2l_color_gr}
\end{figure}

\begin{figure}
\centering
\includegraphics[width=1.0\textwidth]{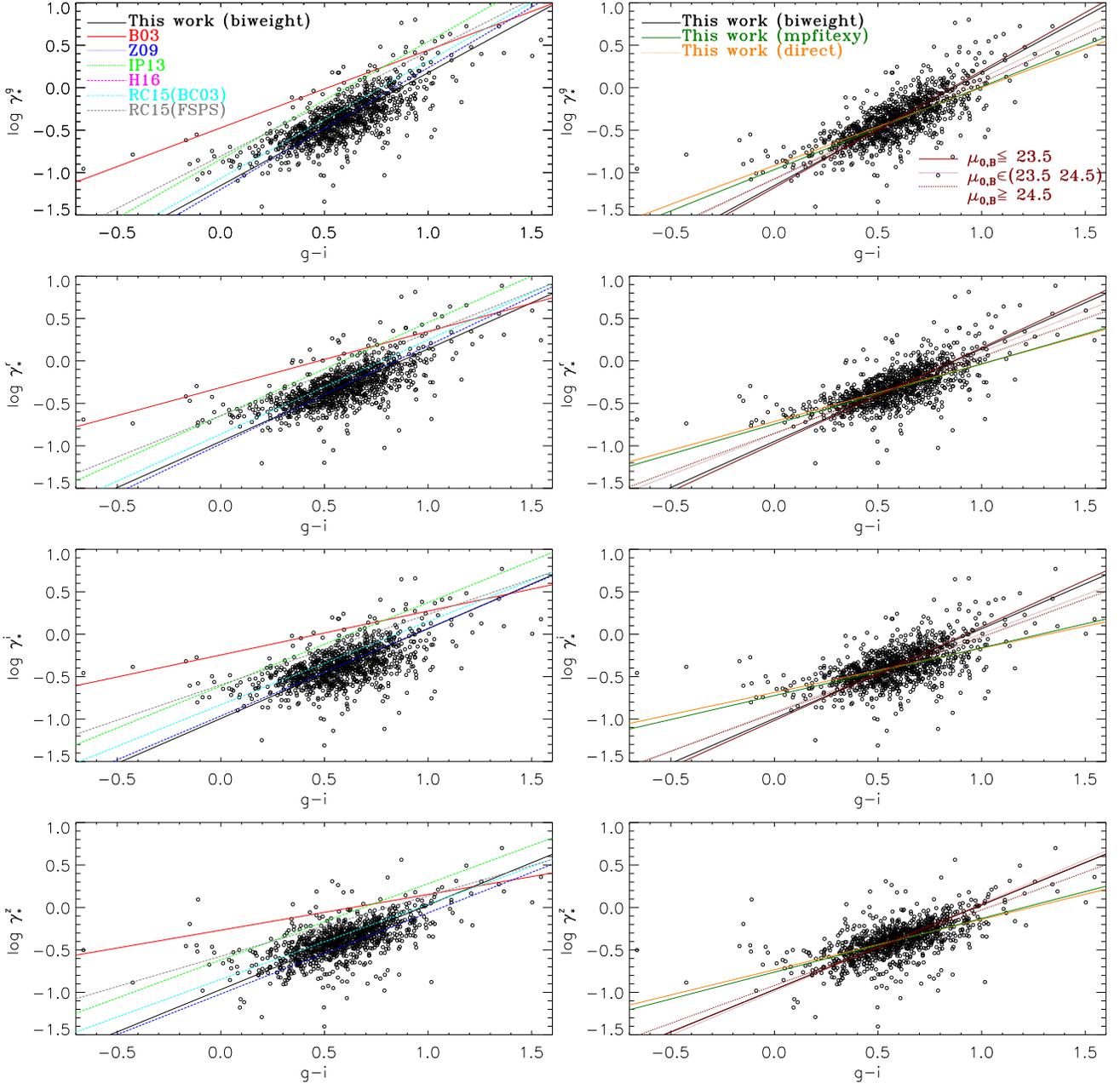}
\caption{Comparison of estimated $g$-,$r$-,$i$- and $z$-band stellar $\gamma^{*}$ ratios
as a function of $g-i$ color for LSBGs in this paper (black circles). In all panels, we show our correlation
by a ``robust" bi-square weighted line fit (solid black line) with the assumption of
a \citet{Chabrier2003} IMF and the CB07 stellar population model.
For comparison, we overplot the correlations of \citet[B03]{Bell2003} as the red line, \citet[Z09]{Zibetti2009} as the blue line, \citet[IP13]{Into2013} as the green line, 
\citet[H16]{Herrmann2016} as the magenta line, \citet{Roediger2015} based on the BC03 model (RC15(BC03)) as the cyan line, and \citet{Roediger2015}
based on the FSPS model (RC15(FSPS)) as the grey line in each panel. All the CMLRs from literature
shown here are under or have been corrected under the \citet{Chabrier2003} IMF.
We note that the color - M/L correlations are in the form of log$_{10} (M/L)$ = $a_{\lambda}$ + $b_{\lambda}\times$color.}
\label{fig:m2l_color_gi}
\end{figure}

\begin{figure}
\centering
\includegraphics[width=1.0\textwidth]{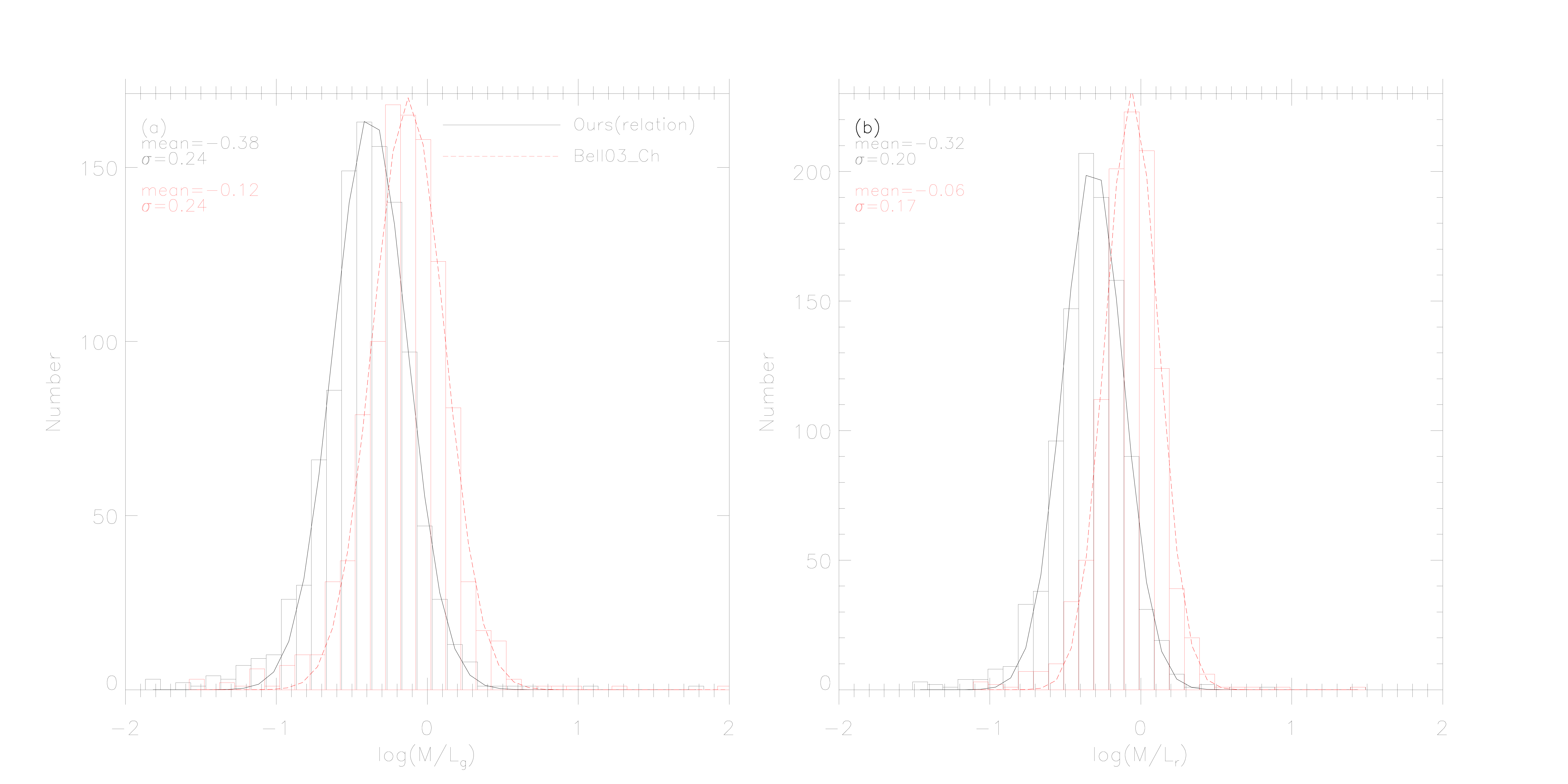}
\caption{Comparisons of $g$-based (left) and $r$-based (right)
$\gamma^{*}$s predicted from Bell03 color-M/L correlations (converted to the \citet[]{Chabrier2003} IMF; red)
with those predicted from our correlations (black) for our LSBG sample.
In each panel, each histogram is fitted by a Gaussian profile to derive a mean and $\sigma$ 
of the $\gamma^{*}$, showing that 
the B03 correlations systematically overestimates the $\gamma^{*}$ by $\sim$0.26~dex
for our LSBG sample.}
\label{fig:offset_Bell}
\end{figure}

\begin{figure}
\centering
\includegraphics[width=0.9\textwidth]{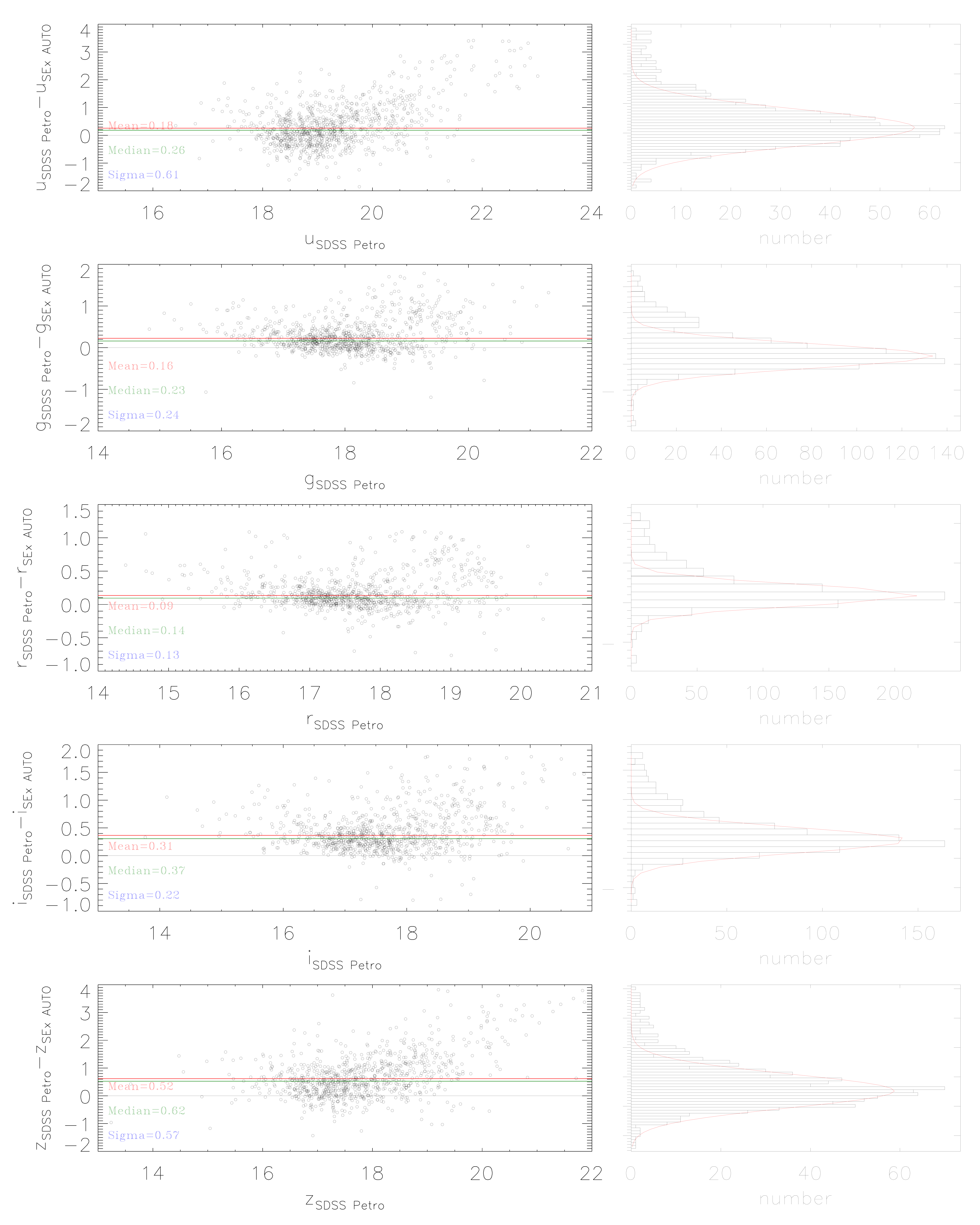}
\caption{Comparison of SEx Kron elliptical aperture magnitude measured with SExtractor
with the SDSS Petro magnitude in the $ugriz$ bands for our LSBG sample. The
panel for each band shows the difference (SEx Kron - SDSS Petro) in magnitude.
In each of the left panels, the black line is the zero difference line. 
Excluding the outlier points outside the $\pm$3$\sigma$,
the robust mean (red line) and median (green line) difference in magnitude and the scatter, $\sigma$, 
are listed in each of the left panels for each band.
The right panels show the histogram distributions of the difference in each band.}
\label{fig:mag_SDSSour}
\end{figure}

\begin{figure}
\centering
\includegraphics[width=1.0\textwidth]{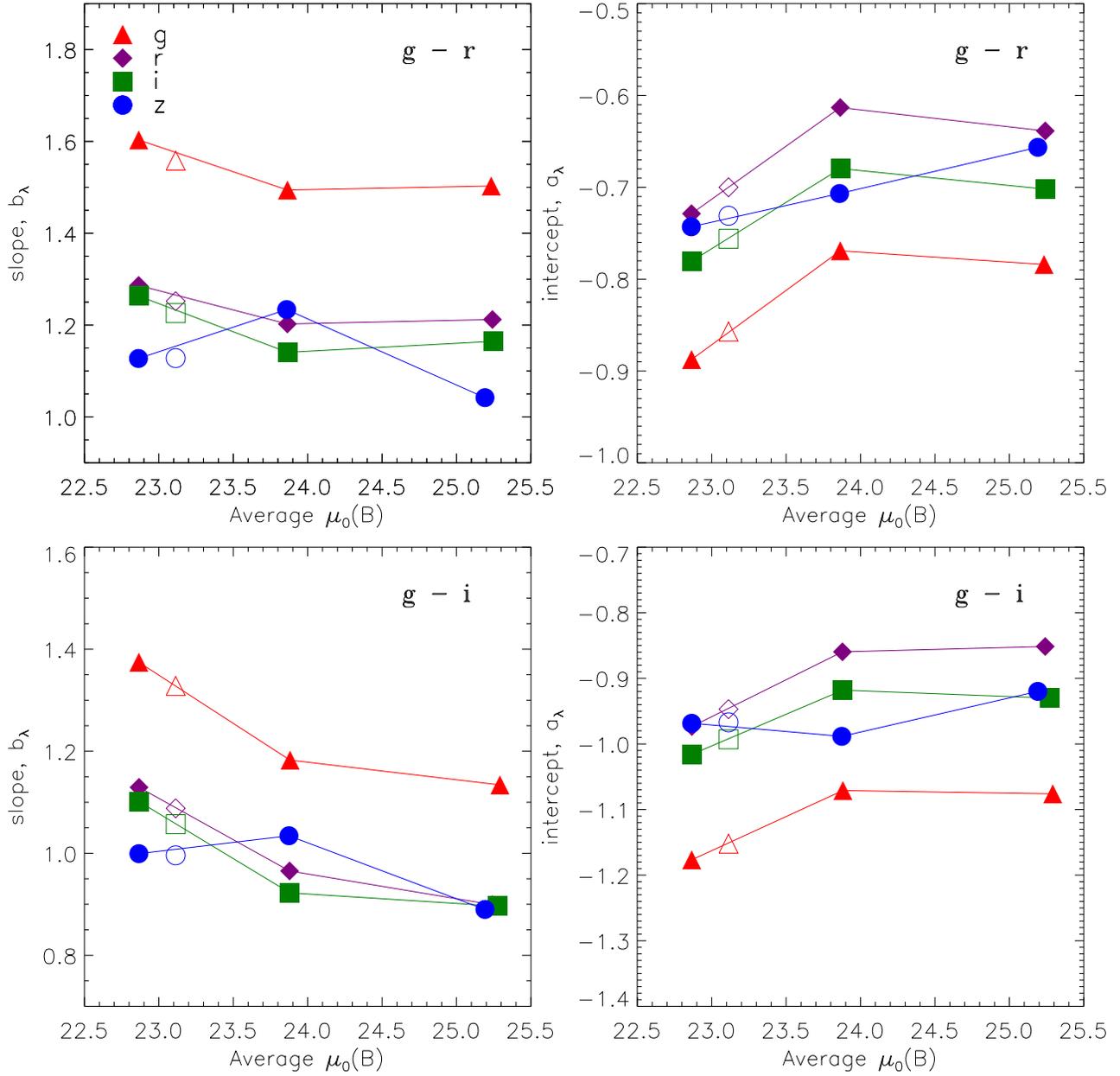}
\caption{MLCR linear fit coefficients determined for three different central surface brightness bins
where log$_{10} (M/L)$ = $a_{\lambda}$ + $b_{\lambda}\times$color.}
\label{fig:slope_mu}
\end{figure}



\end{document}